\documentclass[default,pdflatex]{sn-jnl}
\usepackage[utf8]{inputenc}             
\usepackage[english]{babel}             
\usepackage{duran_emse_twincode_2022}   

\jyear{2023}

\begin{document}

\title[\SHORTTITLE]{\TITLE}

\author*[1,2]{\fnm{Amador} \sur{Durán}}\email{amador@us.es}
\author[1,2]{\fnm{Pablo} \sur{Fernández}}\email{pablofm@us.es}
\author[1,2]{\fnm{Beatriz} \sur{Bernárdez}}\email{beat@us.es}
\author[3]{\fnm{Nathaniel} \sur{Weinman}}\email{nweinman@berkeley.edu}
\author[3]{\fnm{Asl\i{}han} \sur{Akal\i{}n}}\email{asliakalin@berkeley.edu}
\author[3]{\fnm{Armando} \sur{Fox}}\email{fox@berkele.edu}

\affil[1]{%
  \orgdiv{\ITRESUS}, %
  \orgname{\US}, %
  \orgaddress{%
    \city{\Sevilla}, %
    \country{\Spain}%
  }%
}

\affil[2]{%
  \orgdiv{\SCORE}, %
  \orgname{\US}, %
  \orgaddress{%
    \city{\Sevilla}, %
    \country{\Spain}%
  }%
}

\affil[3]{%
  \orgdiv{\ACE}, %
  \orgname{\UCB}, %
  \orgaddress{%
    \city{Berkeley}, %
    \country{\USA}%
  }%
}




\abstract{%

\noindent\textbf{Context}. 
Women have historically been underrepresented in \SE, due in part to an unwelcoming climate pervaded by the widely-held gender bias that men outperform women at programming. \PP is both widely used in industry and has been shown to increase student interest in \SE, particularly among women; but if those same gender biases are also present in \pp, its potential for attracting women to the field could be thwarted. %
\noindent\textbf{Objective.} 
We aim to explore the effects of gender bias in \pp. Specifically, in a remote setting in which students cannot directly observe the gender of their peers, we study whether the perceived productivity, perceived technical competency of the partner, and collaboration/interaction behaviors of \SE students differ depending on the perceived gender of their remote 
partner. To our knowledge, this is the first study specifically focusing on the impact of gender stereotypes and bias \emph{within} pairs in \pp. %
\noindent\textbf{Method}. 
We have developed an online pair-programming platform (\twincode) that provides a collaborative editing window and a chat pane, both of which are heavily instrumented. %
Students in the control group had no information about their partner's gender, whereas students in the treatment group could see a gendered avatar representing the other participant as a man or as a woman. The gender of the avatar was swapped between programming tasks to analyze 45 variables related to the collaborative coding behavior, chat utterances, and questionnaire responses of 46 pairs in the original study at the \us, and 23 pairs in the external replication at the \UCB. %
\noindent\textbf{Results}. 
We did not observe any statistically significant effect of the gender bias treatment, nor any interaction between the perceived partner's gender and subject's gender, in any of the 45 response variables measured in the original study. %
In the external replication, we observed statistically significant effects with moderate to large sizes in four of the 45 dependent variables within the experimental group, comparing how subjects acted when their partners were represented as a man or a woman. %
%
%
\noindent\textbf{Conclusions}. 
The results in the original study do not show any clear effect of gender bias in remote \pp among current \SE students. %
In the external replication, it seems that students delete more source code characters when they have a woman partner, and communicate using more informal utterances, reflections and yes/no questions when they have a man partner, although these results must be considered carefully because of the small number of subjects in the replication, %
and because when false discovery rate adjustments are applied, only the result about informal utterances remains significant. %
In any case, more replications are needed in order to confirm or refute the results in the same and other \SE students populations. %
}


\maketitle



\section{Introduction} \label{sec:introduction}

Besides being widely used in industry, \pp is becoming increasingly common in \SE education because of its demonstrated positive influence on grades, class performance, confidence, productivity, and motivation to stay in \SE and \CS academic majors \citep{dpp-survey-2015}, especially for women, as reported by \cite{werner-et-al}. %

In \pp, two partners work closely together to solve a programming task, in which their ability to engage collaboratively with each other is essential. %
%
%
%
However, these collaborative interactions can be influenced by implicit gender bias \citep{Hofer2015}, which is a widely observed phenomenon even in highly-structured and professional settings, such as those reported by \cite{jarratt-iticse-2019} and \cite{dpp-survey-2015}, and which is based on the assumption that women are less technically competent than men \citep{Martell1996}. %
%
%
Since research in the social sciences indicates that an individual's behavior is clearly affected by the behavior of their peers \citep{Eckles7316}, we aim to explore how and whether 
gender bias affects the \pp experience among \SE students. %

Our study is based on the hypothesis that gender bias will lead to observable differences based on subjects' perceptions of the gender of their \pp partners, i.e. they will score men and women differently on similar tasks, and they will also behave and communicate differently depending on whether they perceive their partner as a man or as a woman, even though their partner remains the same on all tasks. %
Specifically, in a non-colocated, i.e. remote, \pp setting in which peer gender cannot be directly observed, our goal is to identify the potential effects of gender bias by observing student pairs when the perceived gender of one of the peers changes. %

To study our hypothesis, we have applied methodological triangulation \citep{Denzin}, using several methods to collect data and approaching a complex phenomenon like human behavior from more than one standpoint \citep{Cohen}. %
In our case, three different data sources have been used: (1) questionnaires to measure changes in subjects' perceptions, (2) data collected automatically during the \pp tasks to measure behavioral changes, and (3) data produced by several experimenters analyzing the message interchange during the \pp tasks to measure changes in communication. %

\label{sec:research_questions}

Assuming a remote \pp setting, which has been proved to have similar results than co-located \pp as reported by \cite{stotts-2003-virtual} and \cite{al-2016-effectiveness}, our research questions with respect to subjects' perceptions are the following: %

\begin{description}

\item[\textbf{\RQ{1}}] 
Does gender bias affect perceived productivity compared to solo programming? That is, do perceived differences between in-pair and solo productivity depend on the perceived partner's gender?\newline

\item[\textbf{\RQ{2}}] 
Does gender bias affect the partner's perceived technical competency compared to one's own technical competency? That is, do perceived differences between one's own and partners' technical competency depend on the perceived partner's gender?\newline

\item[\textbf{\RQ{3}}]
Does gender bias affect the partner's perceived positive and negative aspects? That is, do perceived positive and negative aspects of their partners depend on the perceived partner's gender?\footnote{This research question, and its associated variables, were added after the presentation of the related registered report at ESEM'2021 \citep{twincode_ESEM2021}. We thought that including an open question could improve the data collection process.}.\newline 

\item[\textbf{\RQ{4}}]
Does gender bias affect how partners' skills are compared? That is, do perceived partners' skills depend on the perceived partner's gender when they 
are compared?

\end{description}

With respect to the subjects' behavior during remote \pp, assuming that gender bias could cause a subject to be more or less proactive on the programming task, or more or less verbose during chatting, 
our research question---based on what we can automatically measure---is the following: %

\begin{description}

\item[\textbf{\RQ{5}}] 
Does gender bias affect the frequencies or relative frequencies with which each partner produces source code additions, source code deletions, successful validations, failed validations, and chat utterances? That is, do these frequencies depend on the perceived partner's gender?

\end{description}

Regarding subjects' communication during remote \pp, we are interested in knowing whether gender bias affects how subjects communicate with their partners, i.e., whether they use a more formal or informal style, and whether they use some types of chat utterances more than others. Our related research questions are the following: %

\begin{description}

\item[\textbf{\RQ{6}}]
Does gender bias affect the relative frequency of formal and informal chat utterances? That is, does the formality of the messages depend on the perceived partner's gender?\newline

\item[\textbf{\RQ{7}}]
Does gender bias affect the frequency or relative frequency of the different types of chat utterances? That is, do the frequencies of the different types of messages depend on the perceived partner's gender?

\end{description}\vspace{-0.5em}


\subsection{The \twincode platform} \label{sec:twincode}

To support our study, we have developed the \twincode remote \pp platform \citep{twincode_sigcse}, which manages %
(i) the registration of students collecting demographic data; %
(ii) the random allocation into experimental and control groups balancing gender proportions, i.e. trying to have the same number of persons of the same gender in both groups; %
(iii) the random allocation into experimental-control pairs; %
(iv) the random assignment of programming exercises to individual subjects and pairs; %
(v) the swapping of gendered avatars between \pp exercises for those subjects in the experimental group; %
and (vi) the automatic collection of interaction metrics and chat utterances. %

\begin{figure*}
	\centering
  \subfigure[Experimental group --- gendered avatar]{   
    \includegraphics[height=0.45\textheight]{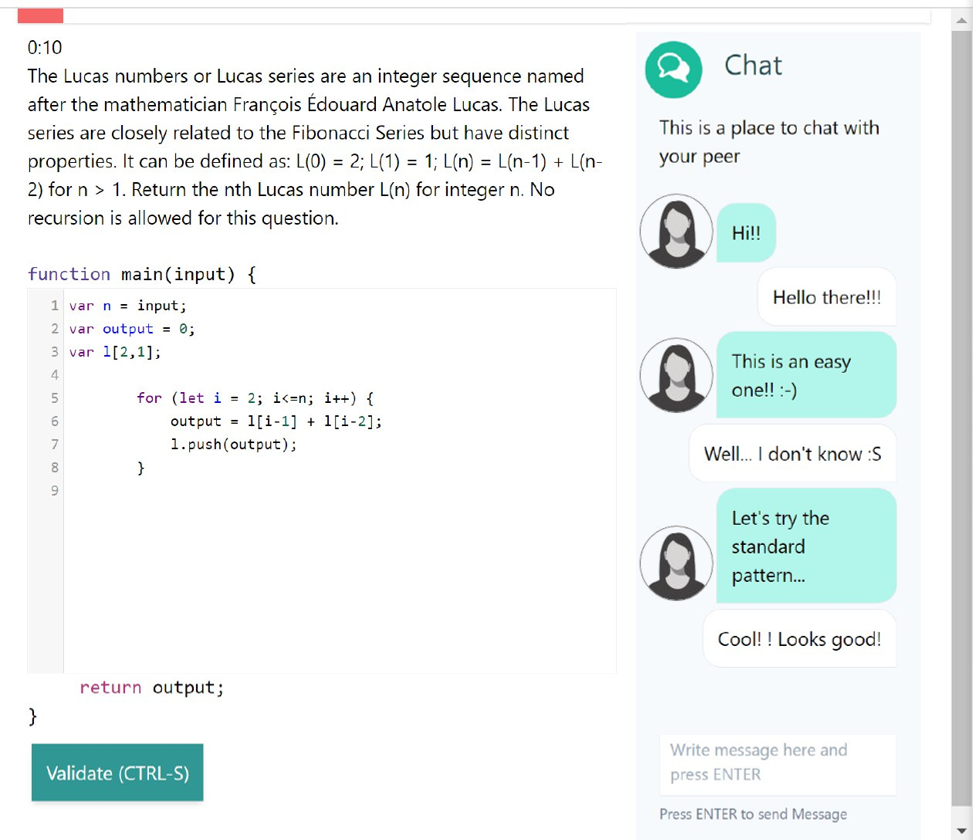}  
    \label{fig:twincode_experimental}
  }
  \subfigure[Control group --- no avatar]{
    \includegraphics[height=0.45\textheight]{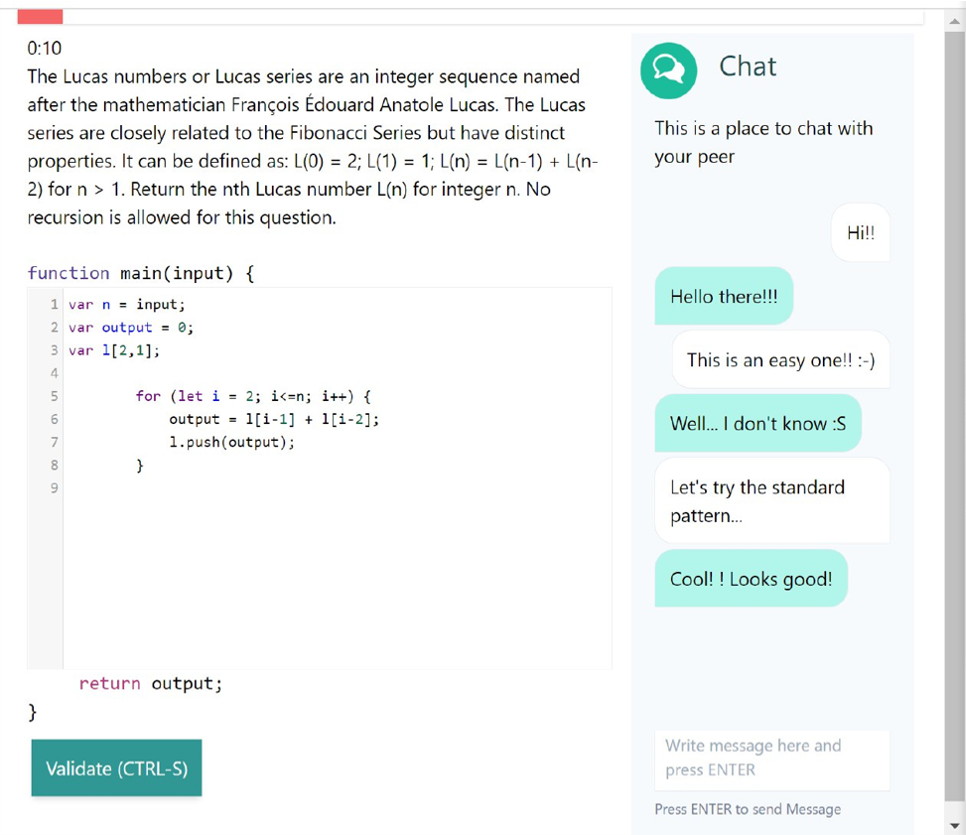}
    \label{fig:twincode_control}
  }
	\caption{\twincode user interface for subjects in the experimental and control groups (original study version)}
	\label{fig:twincode}
\end{figure*}

As shown in \figurename~\ref{fig:twincode}, \twincode offers a source code editor where the students concurrently develop the solution to a proposed programing exercise in Javascript and can validate it against several test cases. %
Note that, to foster communication, only one partner can validate the source code at the same time and see validation results, which should be communicated to the other partner using the chat window, where they are instructed to collaborate to solve the proposed exercises. %
Note also that a gendered avatar is displayed only for the student in the experimental group (see \figurename~\ref{fig:twincode_experimental}) %
but not for the one in the control group (see \figurename~\ref{fig:twincode_control}). %

Experimenters can use \twincode to create new experimental sessions where they can configure, among other aspects, the type, 
number, and duration of the programming exercises, and the instructional messages shown to the students. If needed, they can also develop new programming exercises and their corresponding test cases. %

The \twincode platform is in permanent evolution, and several improvements were incorporated for satisfying some emerging requirements during our study, such as allowing the use of Python as an alternative programming language to Javascript for the programing exercises, changing the images used as gendered avatars (see \figurename~\ref{fig:avatars}), and improving the user interface with instructions and a gendered message in the chat window (see \figurename~\ref{fig:twincode_ucb_experimental} and \ref{fig:twincode_ucb_control} in \appendixname~\ref{sec:app:twincode}).

As a companion tool to \twincode, we have also developed \tagachat, a tool that help experimenters code chat utterances using different sets of tags, as shown in \figurename~\ref{fig:tagachat} in \appendixname~\ref{sec:app:twincode}. %
To assist experimenters during the training stage of the coding, \tagachat automatically computes metrics such as Cohen's \emph{kappa} (for two coders) and Fleiss's \emph{kappa} (for three or more coders) in those dialogs that are being coded by several experimenters to achieve inter-coder reliability assessment \citep{Oconnor2020,Syed2015}\footnote{Although commercial qualitative analysis tools such as MAXQDA (\url{https://www.maxqda.com}) or Atlas.ti (\url{https://atlasti.com}) are available, we decided to develop \tagachat because they are not specifically designed for coding chat utterances, the support for inter-coder reliability metrics is limited, and we prefer to be able to expand its functionality to our future needs and let other researchers use it free of charge.}. %
%

\subsection{Pilot Studies} \label{sec:pilot_studies}


After presenting a very initial approach to our study \citep{Asli2021}, and to get early feedback on (i) the comprehensibility and internal consistency of the scales used in the questionnaires; (ii) the usability and performance of the \twincode platform; and (iii), the applicability of the chat utterance coding based on the one proposed by \cite{Rodriguez2017} and shown in \tablename~\ref{tab:tags}, two pilot studies with a limited number of students were carried out at the \us and \UCB (\ucb) during the 2020--21 academic year. %

As a result, the questionnaires were reorganized into three scales that were assessed for internal consistency (see \appendixname~\ref{sec:app:response_items}), the initial set of chat utterance codes was augmented with formality codes, and the performance and reliability of the \twincode platform was improved. %



\begin{table}[t]
  \centering
  \small
  \renewcommand{\arraystretch}{1.25}  
  \begin{tabularx}{\columnwidth}{cXX}
    \hline\noalign{\smallskip}
    
    \thead{Tag} & \thead{Description} & \thead{Examples} \\
    
    \noalign{\smallskip}\hline\noalign{\smallskip}   
    
    I &
    Informal &
    \emph{LOL! Hahaha!} \\
        
    F &
    Formal &
    All messages except informal \\

		\hline\noalign{\smallskip}        

    S & 
    Statement of information or explanation &
    \emph{We need to create a program for kids to learn math} \\
    
    U & 
    Opinion or indication of uncertainty &
    \emph{Unsure how to add strings together} \\
    
    D & 
    Explicit instruction &
    \emph{Wait put the if back} \\
    
    SU & 
    Polite or indirect instruction &
    \emph{Maybe we can do if user choice = +} \\
    
    ACK & 
    Acknowledgement &
    \emph{Oh ok gotcha} \\
    
    M & 
    Meta-comment or reflection &
    \emph{Hmmm} \\
    
    QYN & 
    Yes/no question &
    \emph{Can the answer be negative?} \\
    
    QWH & 
    Wh- question (who, what, where, when, why, and how) &
    \emph{How do I take in their input?} \\
    
    AYN & 
    Answer to yes/no question &
    \emph{Yea} \\
    
    AWH & 
    Answer to wh-question &
    \emph{The program should be able to generate erroneous questions} \\
    
    FP & 
    Positive task feedback &
    \emph{Oh nice} \\
    
    FNON & 
    Non-positive task feedback &
    \emph{Thats weird} \\

    O & 
    Off-task &
    \emph{Wow its sweet in this room} \\

		\hline\noalign{\smallskip}
  \end{tabularx}
  \caption{Chat utterance tags by \cite{Rodriguez2017} augmented with orthogonal informal/formal tags
  }
  \label{tab:tags}
\end{table}

\subsection{Other Gender Identities} \label{sec:other_genders}

While we recognize that many \SE students may not identify as either men or women, our initial exploration focuses primarily on interactions between students who identify as one of these. 
The potential biases in interactions involving gender-fluid, gender-nonconforming, and nonbinary students is a complex topic deserving its own subsequent study.

\subsection{Structure of the Paper} \label{sec:paper_structure}

The rest of the paper is organized as follows. %
Section \ref{sec:related_work} reviews related work, although to our knowledge, this is the first study specifically focusing on the impact of gender bias \emph{within} pairs in \pp. %
%
%
Sections \ref{sec:original_study} and \ref{sec:first_replication} describe the original study carried out at the \us (December 2021) and its first external replication performed at \ucb (May 2022) respectively. %
%
%
Section \ref{sec:discussion} discusses the two studies and the threats to their experimental validity. %
Finally, Section \ref{sec:conclusions} draws conclusions and proposes future work.


\section{Related Work} \label{sec:related_work}




\begin{sidewaystable}
	\caption{Summary of secondary studies (\SMS or \SLR) in \pp in chronological order}
	\label{tab:related_work_secondary}
	\centering
  \scriptsize
  \renewcommand{\arraystretch}{1.25}
  \scalebox{0.8}{
  \begin{tabularx}{1.2\textheight}{c>{\hsize=.25\hsize}X>{\hsize=.68\hsize}XX}
		\hline\noalign{\smallskip}

		\thead{Reference}             & 
		\thead{Selected Papers}       & 
		\thead{Main Factors Analyzed} & 
		\thead{Conclusions}           \\		

    \noalign{\smallskip}\hline\noalign{\smallskip} 
		
    \cite{salleh-2011} & 
    74 papers selected from 1999 - 2000 period; Controlled, longitudinal, observational studies either investigating factors impacting effectiveness of PP (23\%) or those measuring effectiveness (90\%) via various metrics such as quality (44\%) &
    14 compatibility factors (personality type, actual and perceived skill level, communication skills, self-esteem, gender, ethnicity, learning style, work ethic, time management ability, feel-good factor, confidence level, type of role and type of tasks) and 4 main measures of effectiveness in PP: technical productivity (time spent, knowledge/skill transfer, task performance, code accuracy, number and types of problem, number of solutions to pass test cases), program design and quality (expert opinion, std quality model, code coverage, number of tests passed/failed, LOC, design scores/quality, number of code defects), academic performance (assignment, final, midterm quiz, project and test scores, course grade, course completion rate, retention rate), satisfaction (pair formation, increased knowledge and confidence, positive attitude about collaboration, enjoyment and social interaction) &
    Paired students report higher satisfaction and achieve productivity similar or better than solo students. Implementing PP in the classroom or lab does not lead to any detrimental effect on students’ academic performance. While PP had no significant advantage in improving students’ performance in final exams over solo programming (effect size = 0.16) it was effective in helping students get better scores in their assignments (effect size = 0.67). Personality type, actual and perceived skill levels are investigated the most in PP studies but effects of personality were inconclusive. Pair works well when both students have similar abilities and motivation. Students prefer to pair with someone of similar skills, and students' skill level is the most important factor influencing effectiveness of PP. Most popular metric to measure productivity is time spent on completing tasks. Code quality is another common metric to measure productivity that can be measured as internal, external or general categories. When quality was measured according to academic performance and expert opinion (external), students who pair-programmed produced a better quality program compared to students who programmed alone. However, when the quality of the work produced by pair and solo students was measured using metrics at the internal code level, results were contradictory. \\
    
    \cite{kaur-2021} & 
    68 papers selected via thematic analysis & 
    Comparing PP vs. SP settings, student performance, student attitude and enjoyment &
    PP has mixed effects on students’ performance, but almost universally positive effects on students’ attitudes (i.e. enjoyment) toward programming. Analysis point out the problems such as small sample size in majority of the studies preventing generalization of the results; the lack of context in reported PP experiments obscures validity of results, and that longitudinal analysis of PP experiments with learning tasks of increasing size and complexity is important to build real-world evidence of the practice.\\

    \cite{korber-2021} &
   41 papers published after 2000; &
   Comparing solo vs pair programming, personality, motivation, problem solving, troubleshooting, effectiveness, efficiency, confidence, self esteem, skill level, gender, enjoyment &
   PP positively impacts motivation, self-esteem and confidence of learners. Students report more fun solving the assignments and think PP helps them solve problems faster. Pair programming students reported more effective programming in both visual and text-based languages but earned more achievement points in in the text-based language (python) compared to the visual. PP is effective, but not always efficient. Students who find introductory programming or learning a text-based language especially benefit from the positive effects of pair programming: an appreciative and clear communication regarding possible mistakes and misunderstandings is one of the key factors. Social factors such as gender, personal relationships, effects of successes and failures (attitude), distribution of workload and the influence of partner changes must be considered. \\

    \noalign{\smallskip}\hline 

    \end{tabularx}}
\end{sidewaystable}

Several systematic literature reviews (\SLR{}s), which are summarized in \tablename~\ref{tab:related_work_secondary}, have compiled the empirical research on \pp in higher education, including \citep{dpp-survey-2015}, which is focused on distributed \pp from a teaching perspective. %

The \SLR by \cite{salleh-2010-effects} reveals that the most important factor under study is solo versus \pp in terms of effectiveness, quality of code, and satisfaction while students are programming, concluding that \pp is more effective and satisfactory than solo programming. However, with respect to quality, findings are inconclusive. %

Other \SLR{}s, such as the ones by \cite{hanks-2011-pair}, \cite{Gupta-2021}, and \cite{hawlitschek2022empirical}, show that the focus of the studies is broadened, including factors such as personality, motivation, problem solving, troubleshooting, efficiency, confidence, self-esteem, skill level, gender, or enjoyment but not gender bias. In general, students rate \pp positively compared to solo programming. Nevertheless, \pp is effective but not always efficient, as it may take longer. %

By means of controlled experiments, %
remote and co-located \pp are compared by \cite{stotts-2003-virtual} and \cite{al-2016-effectiveness}, showing similar results. %
In most cases, the analyzed variables are related to performance in terms of time, quality, or code tests passed. Students perceptions have also been analyzed in terms of confidence, satisfaction, motivation, or personality by \cite{salleh-2014-investigating}. %

Regarding primary studies, \tablename~\ref{tab:related_work} summarizes the empirical studies on the influence of gender in \pp, including findings such as %
(i) same-gender pairs are more ``democratic''; %
(ii) women working in pairs were more confident than those working solo; and %
iii) in mixed-gender pairings, women are less confident compared to same-gender pairings, and report no increase in enjoyment for \pp compared to solo programming, an effect that is significantly observed in men \citep{Gupta-2021}. Although such studies reveal that gender seems to be a key factor, none of them study gender bias in \pp. %

Many factors other than gender may affect the outcomes of remote programming sessions \citep{Chaparro2005FactorsAT,Thomas2003}. Previous research on productive pairing looked at factors such as skill levels, autonomy in choosing one’s partner \citep{Xinogalos2017StudentPO}, and different personalities \citep{personality-hannay2010}. %
Nevertheless, the work on gender composition of pairs found conflicting results about whether same-gender or mixed-gender pairings are more effective \citep{choi-bit-2015,genderinpp-choi2013,Hofer2015,kuttal2019}. One possible explanation is that gender correlates with other dimensions that may affect the pairs' collaboration, but these correlations may vary between different environments. For example, women in a class may, on average, have higher skill level than men because they had to face more societal barriers to enter the class. On the other hand, they may, on average, have lower skill level if women with no background are more actively recruited.



\begin{sidewaystable}[p]
	\caption{Summary of primary studies on gender and \pp in chronological order}
	\label{tab:related_work}
	\centering
  \scriptsize
  \renewcommand{\arraystretch}{1.25}
  \scalebox{0.8}{
  \begin{tabularx}{1.2\textheight}{c>{\hsize=.25\hsize}X>{\hsize=.70\hsize}XX}
		\hline\noalign{\smallskip}

		\thead{Reference}       & 
		\thead{Object of study} & 
		\thead{Metrics}         & 
		\thead{Findings}        \\		

    \noalign{\smallskip}\hline\noalign{\smallskip} 
		
    \cite{Katira-icse-2005} & 
    Compatibility of student pair programmers & 
    Web-based peer evaluation survey that required the students to evaluate the contributions of their partner and the perceived pair compatibility & 
    Students are compatible with partners whom they perceive of similar skill. Mixed-gender pairs are less likely to report compatibility. \\
    
      \cite{sfetsos2009experimental} & 
   Effect of personality heterogeneity on PP effectiveness & 
   The Keirsey Temperament Sorter personality test; PP effectiveness is measured by output/performance, communication, velocity, design correctness, passed acceptance tests; pair collaboration-viability is measured by satisfaction, knowledge acquisition and participation. & 
   Heterogeneous personality pairs shows better communication, pair performance and pair collaboration-viability than homogeneous pairs. For heterogeneous pairs, design and code correctness is positively correlated with communication transactions (more communication leads to higher correctness), and satisfaction regarding collaboration, knowledge acquisition and participation was significantly higher.\\
    
    \cite{salleh-2014-investigating} & 
    Personality traits on  PP effectiveness & 
    Five Factor Model (FFM2): Conscientiousness, Neuroticism, and Openness to experience & 
    Only openness has a significant role in differentiating paired students’ academic performance.\\	 
	
    \cite{choi-bit-2015} & 
    PP gender combinations & 
    Productivity, quality of source code, compatibility and communication between pairs &
    Pair compatibility and communication levels significantly vary between the same gender pair type, woman-woman and man-man.\\

    \cite{Gomez2017} & 
    PP gender combinations & 
    Productivity &      
    Similar productivity rates for the three gender pair combinations. Greater variability of productivity rates with mixed gender pairs (man-woman) 
    was observed. \\	
   
    \cite{jarratt-iticse-2019} & 
    PP gender combinations & 
    Weekly attendance, work accomplished during lab and perceived productivity &               
    Students who were randomly assigned a woman partner (rather than a man) attended classes more often, were more confident that the solution was correct, and more confident in the finished product that they developed. However, being assigned a woman partner was also associated with completing a smaller percentage of the assignment. \\
    
   \cite{ying2020understanding} & 
    Effect of structured roles in PP, motivation and stress for men and women &
    Lexical features (number of messages, message length, sentiment), Intrictic Movitation Inventory score (IMI) measuring Interest/Enjoyment, Perceived Competence, Effort/Importance, Pressure/Tension, Perceived Choice, Value/Usefulness, and Relatedness; Self-reported stress, perceived competence, perceived choice, learning gain & 
    No significant differences found between structured vs unstructured PP roles. Women reported significantly higher levels of stress, lower levels of perceived competence in their computing abilities and less perceived choice compared to men during a remote PP activity. Dialogue features significantly correlated with women’s reports of stress, perceived competence, or perceived choice. Women tended to feel more relaxed if their partner sent longer messages on average or used more positive language. \\  
    
    \cite{ying2021cs1} & 
    Analyzing the differences between women and men’s awareness of CS gender gap & 
    Survey (which included six questions related to the gender gap in CS), and some follow-up interviews discussing the experiences and perceptions of CS gender gap. & 
    Men were less aware, had milder beliefs and shallow understanding of the gender disparities in computer science. Women were significantly more aware of the gender gap and felt significantly stronger that efforts should be made to reduce the gender gap. Some participants also expressed discomfort at the idea of opportunities for women within CS because they did not think that those were fair; these students would benefit from understanding the idea of equity over equality.\\  
    
    \cite{galdo2022pair} & 
    Young learners in remote compared to co-local PP & 
    Perception and experiences by means of remote collaboration logs, interviews and self evaluation survey  & 
    Students felt successful in remote approach, had positive experiences with collaboration, reported remote PP made them have more autonomic and efficient in navigation compared to co-located PP. Furthermore, students recognized who partner with friends become more confident throughout the learning process (vs non friend partners). \\

    \noalign{\smallskip}\hline 
    
    \end{tabularx}}
\end{sidewaystable}



\section{Original Study (Seville Dec, 2021)} \label{sec:original_study}

In this section, the original study carried out at the \us in December 2021 is reported, including most of the experimental settings which are in common with the external replication performed at the \ucb in May 2022, reported in Section \ref{sec:first_replication}.



\subsection{Participants} \label{sec:participants}

%
%
In the original study carried out at the \us in December 2021, the participants were third-year students of the Degree in \SE enrolled in any of the three groups of the \RE course taught in Spanish\footnote{There is a fourth group of the \RE course which is taught in English and in which the enrolled students are approximately 50\% Spanish and 50\% Erasmus students coming from other countries in the European Union (EU) or from non-UE countries like Israel or Georgia. They were not invited to participate in the study because their command of Spanish was not good enough to chat with a randomly assigned classmate, who would have undoubtedly identified them as foreign students.}. The final number of valid\footnote{The criteria for considering a subject as valid are strongly dependent on properly performing the experimental tasks, which are described in Section~\ref{sec:execution}. The criteria themselves are specified in Section~\ref{sec:data_analysis}.} subjects was 92, arranged in 46 pairs. Only 9 students could not finish the study because of technical problems during the tasks. %
Considering the 92 valid subjects, 15 identified as \emph{woman} (16.30\%), 1 as \emph{non-binary} (1.09\%), and the rest as \emph{man} (82.61\%) during the registration process. %

Note that, although the percentage of women is low, it is above the average percentage in the Degree in \SE at the \us, which unfortunately is close to 11\% according to the last academic year official statistics \citep{US2020-21}.
Note also that, due to the 9 students dropped by technical reasons, the percentage of women could not be kept the same in the control (6 women, 14.29\%) and experimental (9 women, 19.57\%) groups than in the sample (16.30\%), which was our initial intention. 





\subsection{Experiment Execution} \label{sec:execution}





Some weeks before experiment execution, in order to recruit participants, the students enrolled in the three groups of the \RE course taught in Spanish were motivated to voluntarily participate in the study as an interesting experience in remote \pp, but without mentioning neither that the main goal was to study the potential effect of gender bias, nor they were going to be paired with the same classmate during all the study. %
We also remarked that for the purpose of the study, they must remain anonymous to their partners, so they must neither mention nor ask any personal information, thus not discovering that their partners were always the same person. %
After providing all that information, including that the participation in the study counted for a 5\% bonus on their grades to prevent dropout, the interested students registered in the \twincode platform providing some demographic data and accepting the participation conditions.




\begin{figure}
	\centering
	\includegraphics[scale=0.4]{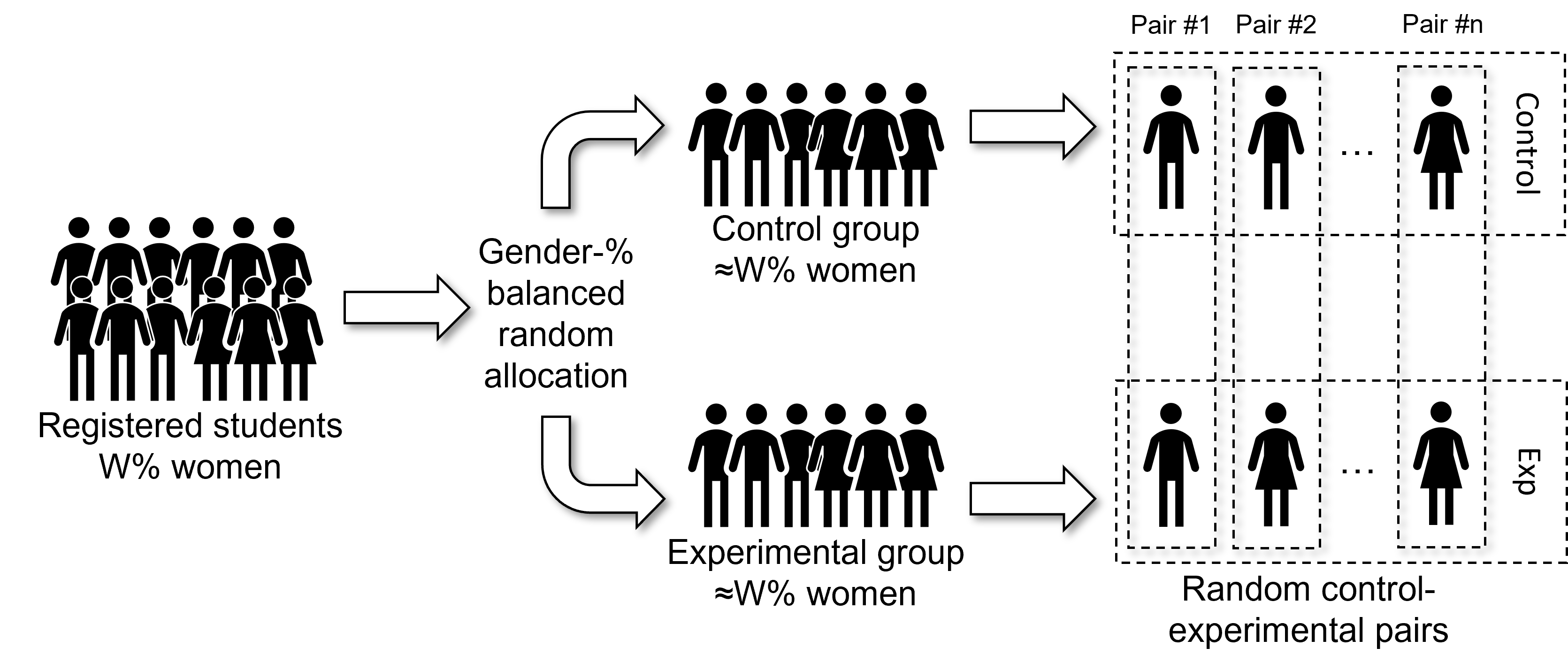}
  \vspace{-1.0em}
	\caption{Experimental process (subject allocation to groups)}
	\label{fig:allocation}
  \vspace{2.0em}
	\includegraphics[scale=0.4]{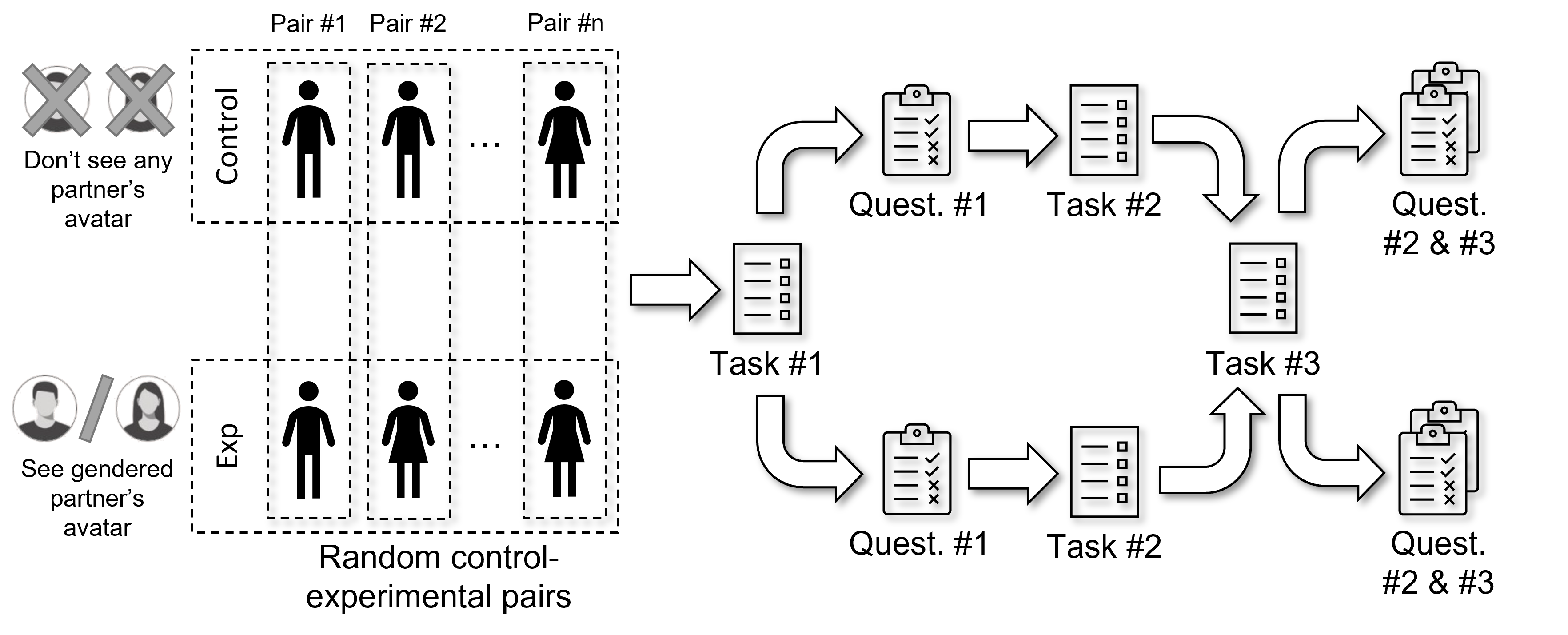}
  \vspace{-0.5em}
	\caption{Experimental process (tasks)}
	\label{fig:tasks}
\end{figure}

\begin{figure}[t]
	\centering
		\includegraphics[width=0.85\textwidth]{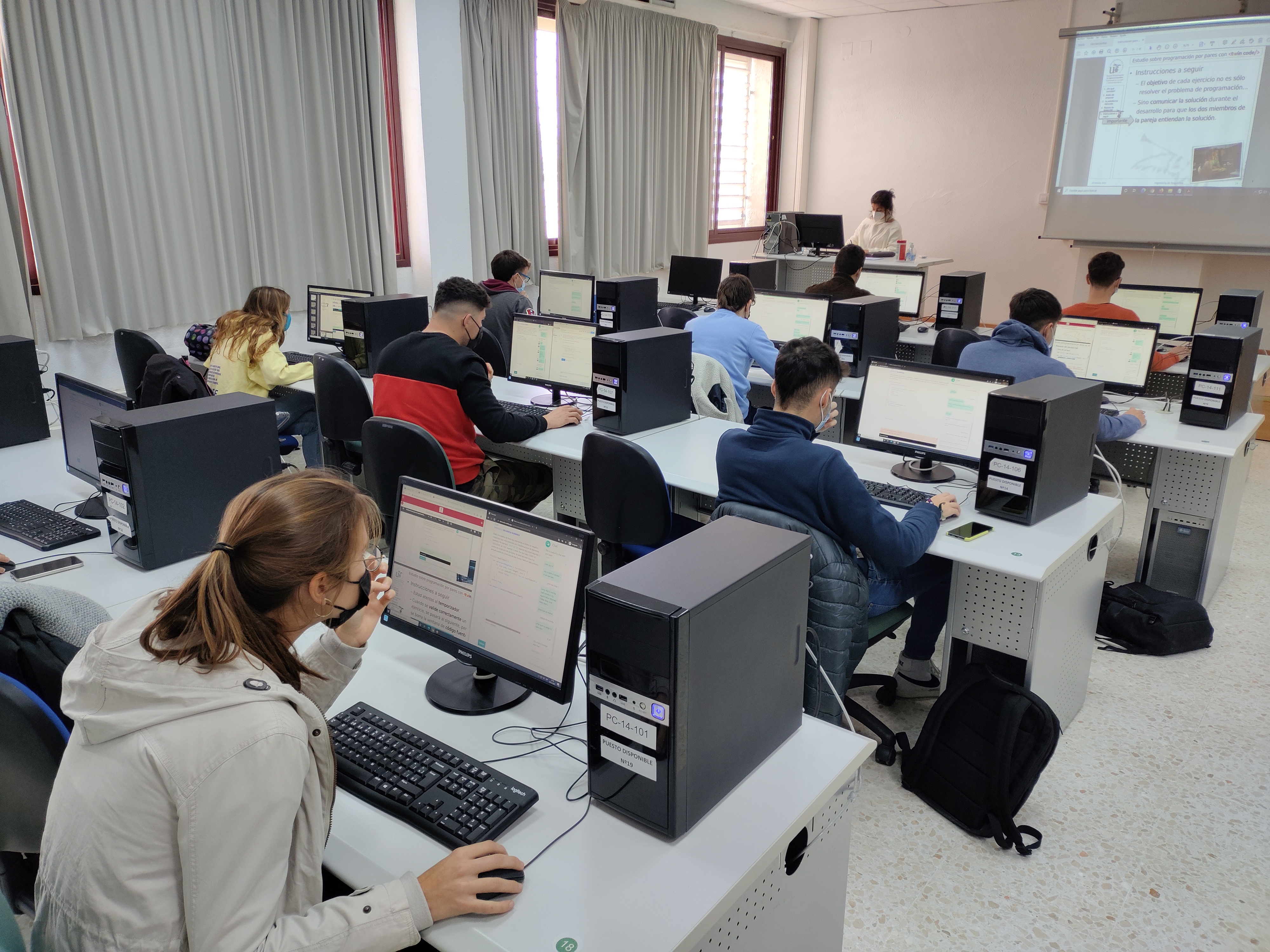}
	\caption{Experiment execution at University of Seville, Dec 2021}
	\label{fig:IMG_20211210_131842}
\end{figure}

The experiment execution, which is graphically represented in \figurename~\ref{fig:allocation} and \ref{fig:tasks}, took place the same day for the three groups of students of the course during their laboratory sessions, as shown in \figurename~\ref{fig:IMG_20211210_131842}\footnote{By the time the experiment was carried out, COVID-19 restrictions in force in Andalucía allowed students to be in the same classroom but wearing masks.}. %

All registered students logged into the \twincode platform, which automatically allocated them into the control and experimental groups balancing the proportion of women in each group as much as possible. %
Once all the students were allocated to groups, they were randomly allocated into control-experimental pairs by the platform (see \figurename~\ref{fig:allocation}). 

After subject allocation, the pairs were presented a programming exercise that they had to solve collaboratively using \twincode (labeled as \variable{Task\#1} in \figurename~\ref{fig:tasks}). %
They were given 10 minutes to solve a first exercise and another 10 minutes to solve a second exercise, thus a total time of 20 minutes. After the first 10-minute period, the second exercise was presented independently of whether the first one was finished successfully or not. Both exercises were randomly selected from a pool of exercises of similar complexity. %
During this programming exercise in pairs, subjects in the control group received no information about the gender of their partners, whereas subjects in the experimental group could see their partners as having a clearly gendered avatar randomly selected by the platform (see \figurename~\ref{fig:twincode}). %
%
%
At the end of the 20-minute period, they were asked to individually fill in a questionnaire (labeled as \variable{Quest.\#1} in \figurename~\ref{fig:tasks}) about the perceived productivity compared to solo programming, the perceived partner's technical competency compared to their own, and about the partner's positive and negative aspects. They were given 10 minutes to fill in the questionnaire.

After filling the first questionnaire, the students were presented another programming exercise to be solved individually in 10 minutes (labeled as \variable{Task\#2} in \figurename~\ref{fig:tasks}). In the case they finished earlier, another exercise of similar complexity was randomly presented. %
The main purpose of this individual task was to make students forget about their first partners, i.e. their style of writing chat utterances or source code, so they did not recognize them in the second in-pair task.

After the individual task, pairs were presented again a new collaborative programming exercise that they must solve in similar conditions to the exercise in \variable{Task\#1}. In this second in-pair exercise, the gendered avatar 
was swapped with respect to the first exercise for the subjects in the experimental group. For those in the control group, they continued to receive no information on their partners' genders. %
%
Note that pairs were kept the same in order to reduce the variability due to the subjects themselves, which could possibly have had a confounding effect in case of a new pair allocation for \variable{Task\#3} (see Section~\ref{sec:confounding_skills} for details).

Once \variable{Task\#3} was finished, students were asked to fill a questionnaire (labeled as \variable{Quest.\#2} in \figurename~\ref{fig:tasks}) with the same questions than the one they filled after \variable{Task\#1} but referred to the second partner, and another questionnaire (labeled as \variable{Quest.\#3} in \figurename~\ref{fig:tasks}) comparing the skills of the first and second partners and whether they remembered the gendered avatars of their partners or not. They were given 15 minutes for responding both questionnaires.

Finally, they were informed about the actual purpose of the study. 
At that point, they were allowed to withdraw their data if they wished, although none of them opted for doing so. 






\subsection{Factors (Independent Variables)} \label{sec:factors}\label{sec:independent_variables}

The four factors, i.e., independent variables, in both the original experiment and the replication are following. 


\begin{description}

\item[\textbf{\variable{group}}]
nominal factor representing the group (\level{experimental} or \level{control}) subjects were randomly allocated to.\newline

\item[\textbf{\variable{time}}] 
nominal factor representing the moment (\level{t$_1$} and \level{t$_2$}) in which the first and second in-pair tasks were performed by the subjects.\newline

\item[\textbf{\variable{ipgender}}] 
nominal factor representing the induced partner's binary gender (\level{man} or \level{woman} for the experimental group, and \level{none} for the control group) during the in-pair tasks. %
%
\newline

\item[\textbf{\variable{gender}}] 
nominal factor representing subject's gender, which may be \level{man}, \level{woman}, or any other option as freely expressed in the demographic form during registration.

\end{description}


\subsection{Response Variables (Dependent Variables)} \label{sec:response_variables}\label{sec:dependent_variables}


The response variables, i.e., dependent variables, in both studies are described below, organized according to the corresponding three data sources---questionnaires, \twincode platform, and chat utterance coding. %

\subsubsection{Perceived Variables (Questionnaires)} \label{sec:perceived_variables}

\begin{figure}
	\centering
		\includegraphics[scale=0.50]{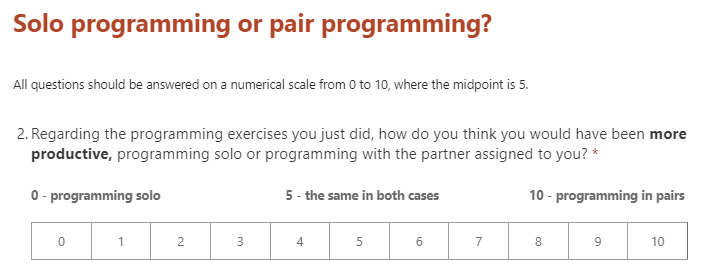}
	\caption{First response item for \variable{pp} variable in questionnaires \#1 \& \#2 as presented to the subjects}
	\label{fig:questionnaire_pp}
\end{figure}

The response variables measuring subjects' perception are mainly scales composed by four or more 0--10 linear numerical response items and they are computed as the average of their corresponding items. %
Following the recommendations by \cite{Versta}, the 0--10 items are labeled not only in the first and last points, but also in the midpoint (see \figurename~\ref{fig:questionnaire_pp}). %
They are described below. 

\begin{figure}
	\centering
		\includegraphics[scale=0.45]{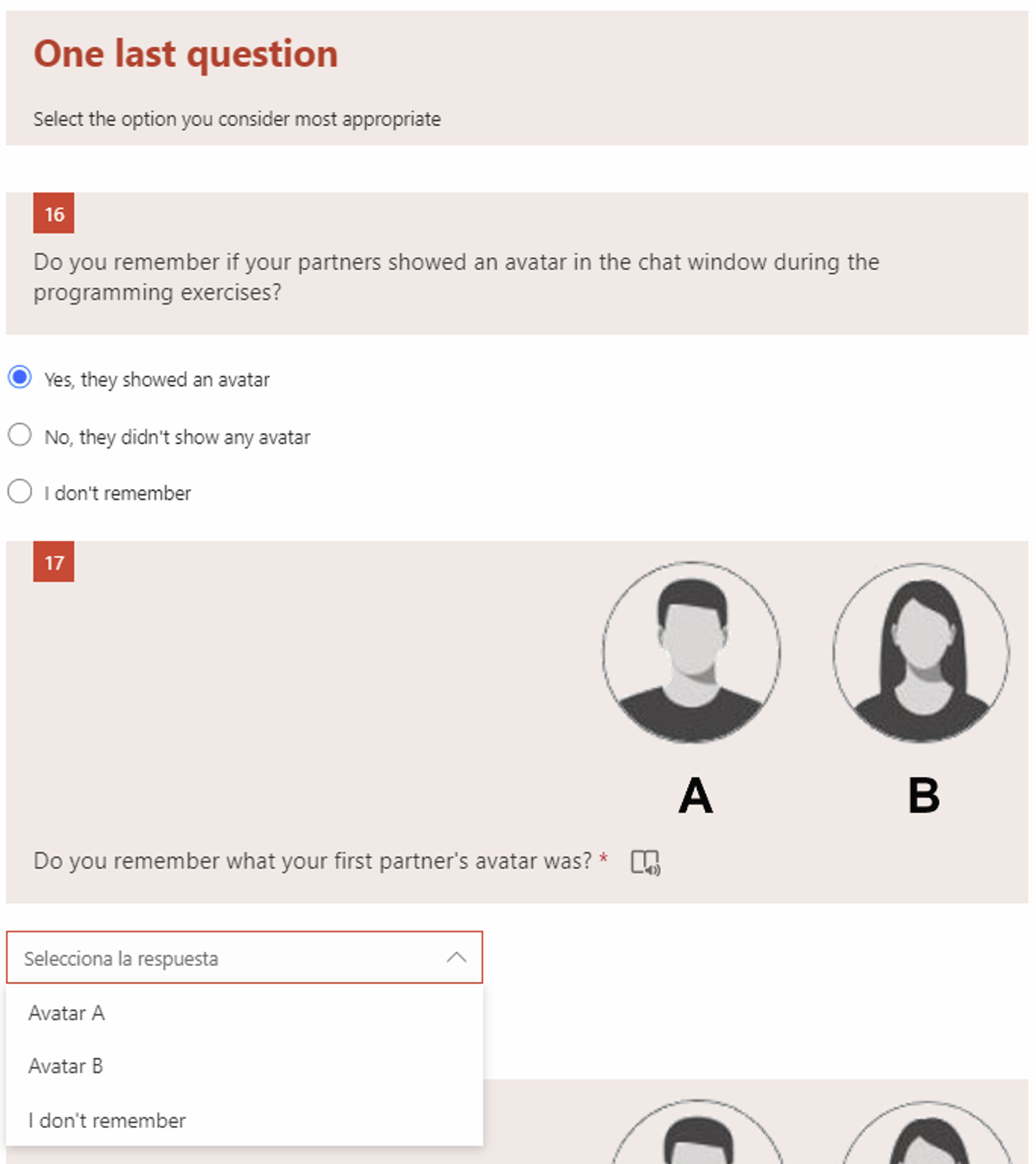}
	\caption{Section in questionnaire \#3 for partner's perceived gender (\variable{ppgender}) variable}
	\label{fig:questionnaire_avatar}
\end{figure}

\begin{description}

\item[\textbf{\variable{pp}}] 
interval variable composed of four 0--10 numerical response items (\variable{pp$_{1\ldots4}$}) measuring the subject's own \textbf{\emph{p}}erceived \textbf{\emph{p}}roductivity during each \pp task compared to solo programming (see \RQ{1}). %
Low values correspond to better solo programming productivity %
whereas high values correspond to better pair programming productivity (see \figurename~\ref{fig:questionnaire_pp} for an example of a response item and Section \ref{sec:app:pp} in the \appendixname\ for all the response items in the scale). %
\newline

\item[\textbf{\variable{pptc}}] 
interval variable composed of four 0--10 numerical response items (\variable{pptc$_{1\ldots4}$}) measuring the subject's \textbf{\emph{p}}artner's \textbf{\emph{p}}erceived \textbf{\emph{t}}echnical \textbf{\emph{c}}ompetency compared to their own after each in-pair task (see \RQ{2}). Low values correspond to higher subject's productivity, %
whereas higher values correspond to higher partner's productivity (see Section \ref{sec:app:pptc} in the \appendixname\ for all the response items). %
\newline

\item[\textbf{\variable{ppa}}] 
ratio variable counting the number of \textbf{\emph{p}}artner's \textbf{\emph{p}}ositive \textbf{\emph{a}}spects identified by the subject after each in-pair task (see \RQ{3})\footnote{According to the four scales of measurements introduced by \cite{Stevens}, variables \variable{ppa} and \variable{pna} are defined as ratio variables because they are numerical variables in which zero represents a lack of the attribute (see Section 2.2 in \citep{Navarro2018} for an excellent explanation, or \citep{GraphPad} for a graphical representation). Note that this is not the case for the \variable{pp}, \variable{pptc}, and \variable{cps} interval variables, in which zero usually means ``the same in both cases'' or ``both equally''.}.
This variable is automatically computed from an open question item in which subjects are asked to write the most positive and negative aspects of their partners in the previously performed \pp exercise (see Section~\ref{sec:app:ppna} in the \appendixname). They are instructed to prefix positive aspects with a plus sign (+) and negative ones with a minus sign (-). This variable is the result of automatically counting the number of plus signs in the text of the open question.
\newline

\item[\textbf{\variable{pna}}] 
ratio variable counting the number of \textbf{\emph{p}}artner's \textbf{\emph{n}}egative \textbf{\emph{a}}spects identified by the subject after each in-pair task (see \RQ{3}). In a similar way to the \variable{ppa} variable, this variable is the result of automatically counting the number of minus signs in the text of the aforementioned open question (see also Section~\ref{sec:app:ppna} in the \appendixname).
\newline

\item[\textbf{\variable{ppgender}}] 
nominal variable measuring the \textbf{\emph{p}}erceived \textbf{\emph{p}}artner's \emph{gender} during the in-pair tasks. To measure this variable, subjects are asked in questionnaire \#3 whether they remember if their partners showed some avatars in chat windows or not. %
If the answer is \level{no} or \emph{I don't remember} (\level{idr}), this variable is assigned the \level{none} or \level{idr} levels at \level{t$_1$} and \level{t$_2$}.
%
%
If the answer is \level{yes}, then the subjects are asked for the avatars of the %
first and second partner, having \variable{man}, \variable{woman}, or \level{idr} as options, as shown in \figurename~\ref{fig:questionnaire_avatar}. 
\newline

\item[\textbf{\variable{cps}}] 
interval variable composed of five 0--10 numerical response items (\variable{cps$_{1\ldots5}$}) measuring whether the subject perceived better skills in their first or second partner in the in-pair tasks, i.e., \textbf{\emph{c}}ompared \textbf{\emph{p}}artners' \textbf{\emph{s}}kills (see \RQ{4}). Low values correspond to the first partner, %
whereas high values correspond to the second partner (see Section \ref{sec:app:cps} in the \appendixname\ for all the response items).\newline %

In the case of the experimental group only, this variable is transformed after collection 
in such a way that low values correspond to the partner for whom the induced gender was \variable{man}, and high values to the partner for whom the induced gender was \variable{woman}, in order to analyze whether there is a gender bias in the scoring. %

\end{description}


\subsubsection{Behavior-Related Variables (\twincode Platform)} \label{sec:behavior_variables}

The response variables automatically collected by the \twincode platform and related to the behavior during the in-pair programming exercises (see \RQ{5}) are listed below. %
Every variable \variable{v} represents a frequency, i.e., a count, and its associated relative frequency is computed with respect to the the sum of the frequencies of the two subjects in a pair. %
For example, let us suppose that subjects $i$ and $j$ are the two members of a pair, and \variable{v$_i$} and \variable{v$_j$} are the corresponding values of the \variable{v} variable. %
In this case, the relative frequencies for each subject would be $\variable{v}_{i} \over{\variable{v}_{i} + \variable{v}_{j}}$ and $\variable{v}_{j} \over {\variable{v}_{i} + \variable{v}_{j}}$, respectively. 


\begin{description}

\item[\textbf{\variable{sca}}~/~\textbf{\variable{sca\_rf}}]
Ratio scale variables representing the count and relative frequency of characters added by a subject to the source code window during an in-pair task (\textbf{\emph{s}}ource \textbf{\emph{c}}ode \textbf{\emph{a}}dditions).\newline

\item[\textbf{\variable{scd}}~/~\textbf{\variable{scd\_rf}}]
Ratio scale variables representing the count and relative frequency of characters deleted by a subject from the source code window during an in-pair task. (\textbf{\emph{s}}ource \textbf{\emph{c}}ode \textbf{\emph{d}}eletions).\newline

\item[\textbf{\variable{okv}}~/~\textbf{\variable{okv\_rf}}]
Ratio scale variables representing the count and relative frequency of successful (\textbf{\emph{ok}}) \textbf{\emph{v}}alidations of the source code performed by a subject during an in-pair task.\newline

\item[\textbf{\variable{kov}}~/~\textbf{\variable{kov\_rf}}]
Ratio scale variables representing the count and relative frequency of unsuccessful (\textbf{\emph{ko}}) \textbf{\emph{v}}alidations of the source code performed by a subject during an in-pair task. \newline

\item[\textbf{\variable{dm}}~/~\textbf{\variable{dm\_rf}}]
Ratio scale variables representing the count and relative frequency of \textbf{\emph{d}}ialog \textbf{\emph{m}}essages (chat utterances) sent by a subject during an in-pair task.

\end{description}

\subsubsection{Communication-Related Variables (Utterance Tagging)} \label{sec:tagging_variables}


The chat utterances registered in the \twincode platform during the in-pair tasks were manually tagged according to two orthogonal dimensions. %
The first dimension uses the 13 tags (from \level{S} to \level{O} in \tablename~\ref{tab:tags}) proposed by \cite{Rodriguez2017}. %
%
The second dimension classifies each message as \level{formal} or \level{informal}, considering as formal the usual way in which a university student would communicate textually to a professor and informal otherwise. %

For the tagging process, we followed a process inspired by the work of \cite{Oconnor2020}, in which two researchers each tagged 60\% of the data, covering all dialogue messages. The overlapping subset of 20\%, which was used for the initial training, established the inter-coder reliability using Cohen’s \emph{kappa}, which was $\kappa$ = 0.796 for the formal/informal tags, and $\kappa$ = 0.754 for \citeauthor{Rodriguez2017} tags, both indicating \emph{substantial} agreement and sufficient reliability for further coding according to \cite{Syed2015}.

The response variables related to the manual tagging of the chat utterances (see \RQ{6} and \RQ{7}) correspond to the  tags in \tablename~\ref{tab:tags} and are listed below. %
%
Every variable represents a frequency, i.e., a count, and its associated relative frequency is computed with respect to the number of chat utterances generated by the subject during an in-pair task, which is defined by the \variable{dm} variable specified in previous section. %



\newcommand{\tagname}[1]{\item[\textbf{\variable{#1}}~/~\textbf{\variable{#1\_rf}}]}
\newcommand{\tagvar}[1]{Ratio scale variables representing the absolute and relative frequency of #1 messages generated by a subject during an in-pair task.}


\begin{description}

\tagname{i} \tagvar{\textbf{\emph{i}}nformal} \newline

\tagname{f} \tagvar{\textbf{\emph{f}}ormal} \newline

\tagname{s} \tagvar{\textbf{\emph{s}}tatement of information or explanation} \newline

\tagname{u} \tagvar{opinion or indication of \textbf{\emph{u}}ncertainty} \newline

\tagname{d} \tagvar{explicit or \textbf{\emph{d}}irect instruction} \newline

\tagname{su} \tagvar{polite or indirect instruction or \textbf{\emph{su}}ggestion} \newline

\tagname{ack} \tagvar{\textbf{\emph{ack}}nowledgment} \newline

\tagname{m} \tagvar{\textbf{\emph{m}}eta--comment or reflection} \newline

\tagname{qyn} \tagvar{\textbf{\emph{y}}es/\textbf{\emph{n}}o \textbf{\emph{q}}uestion} \newline

\tagname{qwh} \tagvar{\textbf{\emph{wh}}- \textbf{\emph{q}}uestion (who, what, where, when, why, and how)} \newline

\tagname{ayn} \tagvar{\textbf{\emph{a}}nswer to \textbf{\emph{y}}es/\textbf{\emph{n}}o question} \newline

\tagname{awh} \tagvar{\textbf{\emph{a}}nswer to \textbf{\emph{wh}}- question} \newline

\tagname{fp} \tagvar{\textbf{\emph{p}}ositive task \textbf{\emph{f}}eedback} \newline

\tagname{fnon} \tagvar{\textbf{\emph{non}}--positive task \textbf{\emph{f}}eedback} \newline

\tagname{o} \tagvar{\textbf{\emph{o}}ff--task}

\end{description}

\subsection{Confounding Variables} \label{sec:confounding_variables}

The confounding variables that were controlled during both studies are described below. 


\subsubsection{Subject's technical skills} \label{sec:confounding_skills}

To control the variability caused by each subject on their partner, pairs were kept the same during the entire experiment, although the subjects were not informed about this fact. 
Ideally, this would make the conditions of the two in-pair tasks the same except for the programming exercises (see below) and for the induced gender in the case of the experimental group.

\subsubsection{Programming exercises} \label{sec:confounding_exercises}

In order to avoid potential differences among the programming exercises used during in-pair tasks, they were all of similar complexity and were randomly assigned. %





\subsection{Data Analysis} \label{sec:data_analysis}

The data analysis was performed only for those subjects considered as valid according to the following criteria: %
(i) to have filled in both questionnaires; %
(ii) to have their metrics correctly collected by the \twincode platform; %
(iii) to have been paired with another valid subject; and %
(iv) not to have disclosed their gender or their partner's during the in-pair exercises; %

This resulted in 46 pairs, i.e. 92 valid subjects, with only 9 subjects dropped because of technical problems with their connections to the \twincode platform, as previously mentioned in Section~\ref{sec:participants}. 


\subsubsection{Correlation between Induced and Perceived Gender} \label{sec:correlation}

Before analyzing between and within-group relationships, the correlation of the induced and perceived gender in both groups was analyzed in order to know whether the treatment had been effectively administered to the subjects\footnote{The analysis of the correlation between induced and perceived gender was not included in the registered report originally submitted to ESEM'2021 \citep{twincode_ESEM2021}. We included it thanks to the reviewers' comments, whose suggestion has definitely improved our analysis.}. %

\begin{table}
  \centering
  \small
  \renewcommand{\arraystretch}{1.25}  
  \begin{tabularx}{0.85\textwidth}{cCCCC}
    
    & \multicolumn{4}{c}{\thead{Perceived Gender}} \\
    
    \cline{2-5}\noalign{\smallskip}
    
    \thead{Induced Gender}   & 
    \thead{\variable{man}}   & 
    \thead{\variable{woman}} & 
    \thead{\variable{none}}  & 
    \thead{\variable{idr}}   \\
    
    \noalign{\smallskip}\hline\noalign{\smallskip}   

    \thead[r]{\variable{man}}   & \textbf{28} (60.87\%) &  1 (2.17\%)  &  6 (13.05\%) & 11 (23.91\%) \\
    \thead[r]{\variable{woman}} &  1 (2.17\%)  & \textbf{27} (58.70\%) &  6 (13.05\%) & 12 (26.09\%) \\
    \thead[r]{\variable{none}}  &  0 (0.00\%)  &  0 (0.00\%)  & \textbf{52} (56.52\%) & 40 (43.48\%) \\

		\hline\noalign{\smallskip}
  \end{tabularx}
  \caption{Contingency table for induced partner's gender (\variable{ipgender}) vs. perceived partner's gender (\variable{ppgender})}
  \label{tab:contingency}
\end{table}

For that purpose, the results of the contingency table in \tablename~\ref{tab:contingency} were analyzed observing that the percentage of subjects who were induced to think that their partner was a \level{man} and that effectively remembered they saw a \level{man} avatar was close to 61\%, whereas in the case of \level{woman} avatars the percentage was close to 59\%. %
Although Cramer's V for \tablename~\ref{tab:contingency} showed a \emph{large} effect (0.709) according to \cite{CramersVthresholds}, we decided to exclude from the remaining analyses those subjects in the experimental group for whom the induced and perceived gender did not match, because we considered that the treatment had not been sufficiently effective in their cases\footnote{We applied this strict selection of subjects in the experimental group in a manner consistent with the results of the correlation analysis, considerably reducing the number of subjects, especially in the replication reported in Section~\ref{sec:first_replication}.}. 
On the other hand, we kept those subjects in the control group who did not perceived any gendered avatar or did not remember it, discarding the rest.
As a result, we kept all the subjects in the control group (39 men, 6 women, 1 non-binary) but only 27 (21 men, 6 women) in the experimental group. %

\subsubsection{Between-groups Analysis} \label{sec:between}

In the analysis between the control and experimental groups, for every response variable \variable{v} except for \variable{cps}\footnote{As commented in its description in Section~\ref{sec:perceived_variables}, the \variable{cps} variable is measured only once at the end of the experimental process, since it compares first and second partners' skills.}, we computed the distance between the two in-pair tasks as the absolute value of the difference, i.e. $|$\,\variable{v}(t$_2$) $-$ \variable{v}(t$_1$)\,$|$, since the sign of that difference was not relevant in our case. %
%
In our research hypothesis, this distance should be smaller for the students in the control group, who received no information about their partners' genders i.e. no treatment, than for those in the experimental group who effectively perceived two different partners' genders at \level{t$_1$} and \level{t$_2$}. %
Therefore, for every response variable except for \variable{cps}, we performed a one-tailed unpaired mean difference test between groups, applying a t-test or a  Mann-Whitney U test (also known as Wilcoxon test), depending on the results of the normality assumption tests. %

In the case of the \variable{cps} variable, for the control group we expected the mean to be closer to the middle point (5) between the first and second partner,  as they were unconsciously comparing the skills of the same person. %
For the experimental group, we expected the mean to be skewed towards 0 (partner perceived as a \variable{man}) or 10 (partner perceived as a \variable{woman}) due to the effect of the treatment. %
Therefore, to detect differences between groups for the \variable{cps} response variable, we performed an unpaired two-tailed t-test because data distribution was not significantly different from normal distribution. 

\begin{figure}
	\centering
		\includegraphics[scale=0.385]{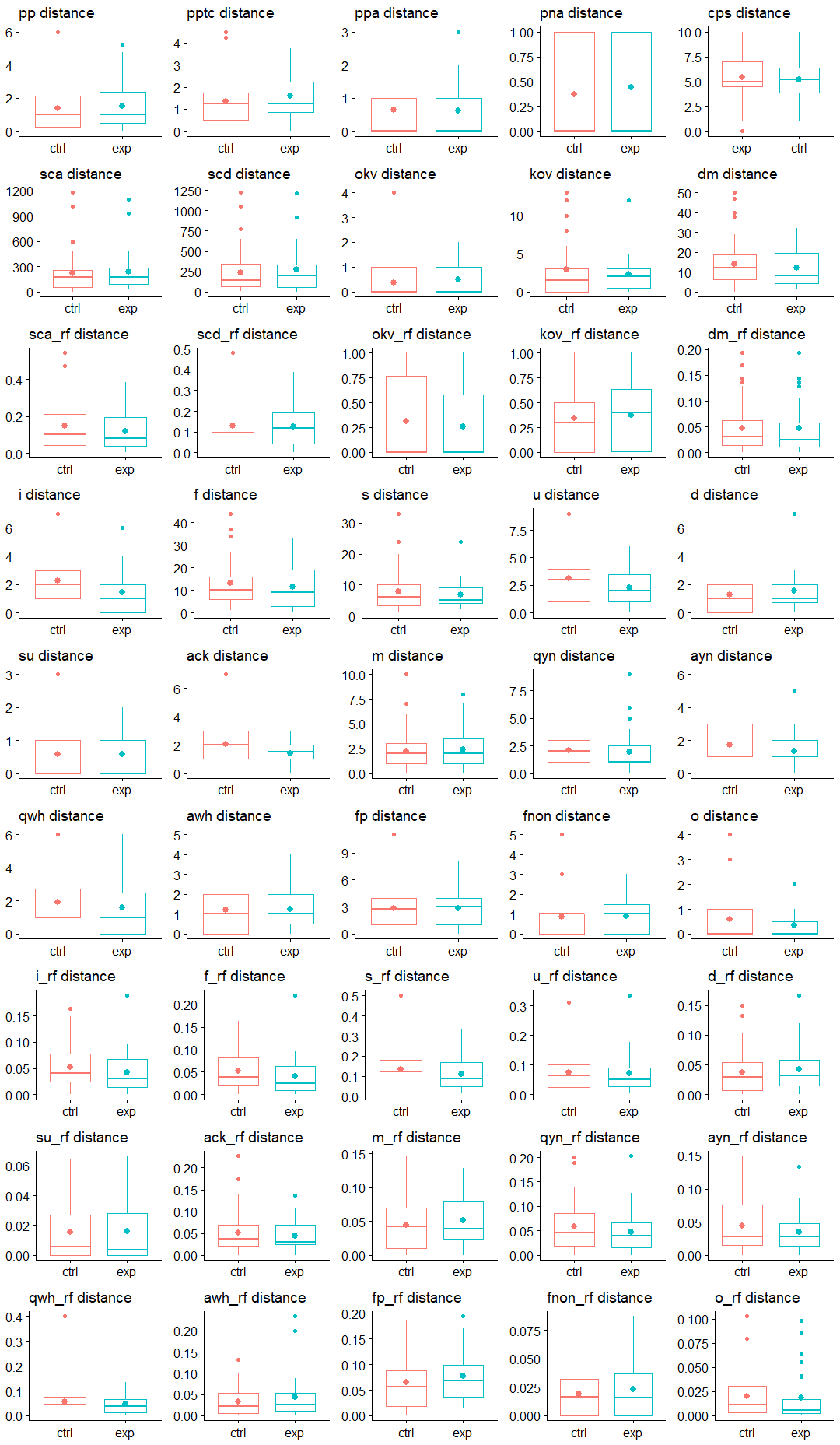}
	\caption{Boxplots of the 45 response variables for between-groups analysis in the original study}
	\label{fig:boxplot_seville_between}
\end{figure}

Contrary to our research hypothesis, no significant differences were observed at $\alpha$=0.05 between the control and experimental groups for any of the 45 response variables described in Section~\ref{sec:dependent_variables}, including \variable{cps}. The corresponding boxplots are depicted in \figurename~\ref{fig:boxplot_seville_between}, where it can be seen that the difference between means---the circles in the boxes---in both groups were very small. %

\subsubsection{Within-groups Analysis} \label{sec:within}

Within the experimental group, we wanted to analyze whether there were differences between the response variables when the same subjects perceived theirs partners as \level{men} or \level{women} according to our research hypothesis. %
We also wanted to study the possible interaction between the perceived partner's gender and the subject's gender. 

For those purposes, we performed a two-sided paired mean difference test for every response variable except for \variable{cps}, using the perceived gender (\variable{ppgender}) as a within-subjects variable, and applying a t-test or a Wilcoxon test depending on the results of the normality assumption tests. %
For studying the interaction, we performed the corresponding mixed-model two-way \ANOVA{}s with the perceived gender (\variable{ppgender}) as a within-subjects variable and the subject's gender (\variable{gender}) as a between-subjects variable. %

For the \variable{cps} variable, which passed the Shapiro-Wilk normality tests, we analyzed whether the subject's gender had any effect when comparing partners perceived as \level{man} or \level{woman} by means of a two-tailed unpaired t-test between groups, using \variable{gender} as a between-subjects variable.

Contrary to our research hypothesis, no significant differences were observed at $\alpha$=0.05 between the two levels of the \variable{ppgender} variable for any of the 44 response variables described in Section~\ref{sec:dependent_variables}. %
None of the 44 \ANOVA tests detected any significant interaction either, and no effect of the subject's gender on the \variable{cps} variable was detected. 

\begin{figure}
	\centering
		\includegraphics[scale=0.385]{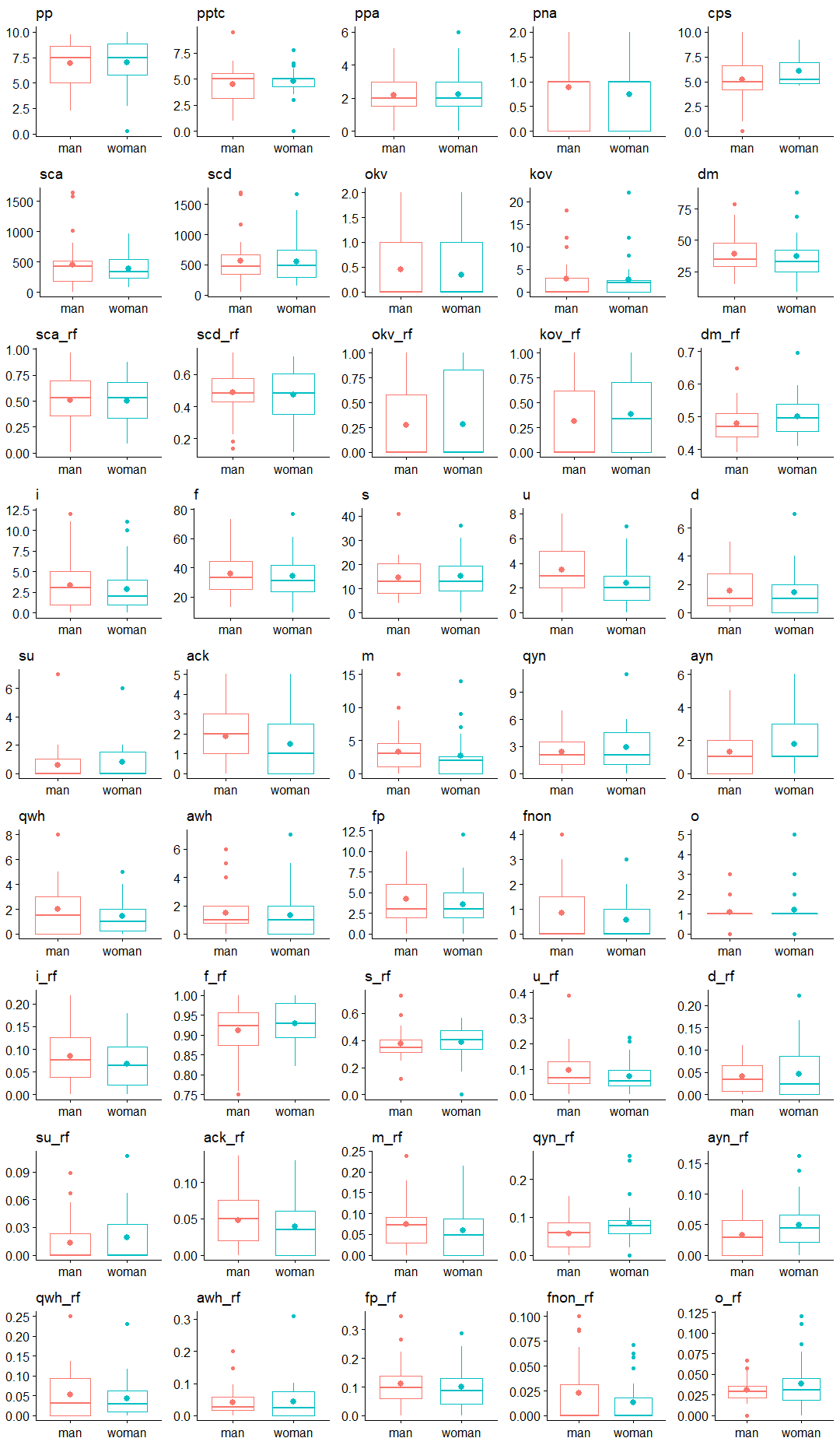}
	\caption{Boxplots of the 45 response variables for within-groups analysis in the original study}
	\label{fig:boxplot_seville_within}
\end{figure}

As depicted in \figurename~\ref{fig:boxplot_seville_within}, the corresponding boxplots show very small differences between means when partners are perceived as \level{men} or \level{women} in the experimental group. %

\section{First Replication (Berkeley May, 2022)} \label{sec:first_replication}

In this section, the first replication carried out at the University of California Berkeley in May 2022 is reported focusing mainly on the changes in the participants and the experiment execution with respect to the original experiment, since the research questions and variables were the same in both studies. %
For each change, an estimation of their impact on the four types of experimental validity described by \cite{Wohlin} is included, following the recommendations by \cite{Margarita_Computing} about reporting the impact of changes in replications using a 7-point discrete scale from $-3$ to $+3$. 
A summary of the impact of those changes is presented in \tablename~\ref{tab:effects_on_validity}, including the labels of the aforementioned scale in its legend. %

\begin{table}[h]%
  \centering
  \small
  \renewcommand{\arraystretch}{1.25}  
  \begin{tabularx}{\textwidth}{lCCCC}
    
    & \multicolumn{4}{c}{\thead{Effect on experimental validity}} \\
    
    \cline{2-5}\noalign{\smallskip}
    
    \thead[l]{Change description} & 
    \thead{Construct}          & 
    \thead{Internal}           & 
    \thead{External}           & 
    \thead{Conclusion}         \\
    
    \noalign{\smallskip}\hline\noalign{\smallskip}
    

    Third to first year students             &  --  &  --  & $+2$ &  --  \\
    Spanish students to U.S. students        &  --  &  --  & $+2$ &  --  \\
    Higher percentage of women               &  --  &  --  & $+2$ &  --  \\
    Number of subjects reduced               &  --  &  --  &  --  & $-2$ \\ 
    5\% grade bonus to \$15 Amazon gift card &  --  &  --  &  --  &  --  \\ 
    Remote location of subjects              & $+2$ & $-1$ &  --  &  --  \\ 
    Higher number of sessions                &  --  & $-1$ &  --  &  --  \\    
    Reduced time for tasks                   & $-1$ & $-1$ &  --  &  --  \\ 
    Different avatars                        & $-1$ &  --  &  --  &  --  \\
    Blocked exercise assignment              &  --  & $+2$ &  --  &  --  \\ 
    Different programming language           &  --  &  --  &  --  &  --  \\
    
		\noalign{\smallskip}\hline\noalign{\smallskip} 
\end{tabularx}

\begin{tabularx}{\textwidth}{lX}
Legend: & --: it does not affect; $-1/+1$: slightly increases/decreases;\\ 
        & $-2/+2$: moderately increases/decreases; $-3/+3$: substantially increases/decreases
\end{tabularx}

\caption{Estimated effects on experimental validity of the changes introduced in the replication}
\label{tab:effects_on_validity}
\end{table}

\subsection{Participants} \label{sec:participants_UCB}

In the replication carried out at the \UCB, the participants were mainly first year students enrolled in the CS61A (\emph{The Structure and Interpretation of Computer Programs}) and CS88 (\emph{Computational Structures in Data Science}) courses. %
Applying the same criteria than for the original experiment, the final number of valid subjects was 46, arranged in 23 pairs. %
Only 6 students, i.e. 3 pairs, were excluded from the initial 52 participants. One pair was dropped due to the disclosure of their identities during the \pp tasks; another pair was dropped because one of its partners did not actively participate in the experimental tasks; and the third pair was excluded because they lost their connection to the \twincode platform repeatedly and their metrics could not be properly collected. %
Among the remaining 46 valid subjects, 26 identified as \emph{woman} (56.52\%) and the rest as \emph{man} (43.48\%) during the registration process\footnote{The only student who reported a non-binary gender described as ``who cares'' was one of the 6 excluded students.}. %

Note that, contrary to the original experiment, the percentage of women is above that of men because the CS61A and CS88 introductory courses are taken also by students from other majors, usually with a higher presence of women than in \CS majors, where is around 25\% \citep{UCB2020-21}. %
Note also that despite the 6 dropped subjects, the percentage of women in the control (12 women, 52.17\%) and experimental (14 women, 60.87\%) groups were close to each other. %

From our point of view, this change in the sampled population from third-year Spanish students to first-year U.S. students, and the higher percentage of women, increased external validity, 
but the reduction in 50\% of the number of subjects (46 pairs to 23 pairs) 
reduced conclusion validity. %

\subsection{Experiment Execution} \label{sec:execution_UCB}

The experiment execution at the \UCB followed the same process than that performed at the \US with some changes, which are described in the following sections.

\subsubsection{Bonus for participating in the study}

As commented in Section~\ref{sec:execution}, in the original experiment the participation in the study counted for a 5\% bonus on students' grades in the \RE course they were enrolled in to prevent dropout. %
In the replication, considering that the students were enrolled in two different courses with different professors, they were offered a \$15 Amazon gift card for participating actively in the study instead of a grade bonus which would have been difficult to manage. %
In our opinion, this change did not affect any type of experimental validity. %

\subsubsection{Location of students and number of sessions} \label{sec:location_change}

In the original experiment, the experimental execution took place during one of the laboratory sessions of the \RE course, as shown in \figurename~\ref{fig:IMG_20211210_131842}. The three groups of the course had the laboratory sessions the same day at different hours, with 30 students per session on average. %
In the replication, the students performed the experimental tasks remotely, coordinated by one of the experimenters using Zoom. There were four sessions that took place during a week with 10 students per session on average. %

We think that this change 
increased construct validity with respect to the original study, since the setting was strictly remote rather than being co-located in a laboratory room, but it also decreased internal validity because of the lack of control of the subject's environment, in which interactions with a third person, interruptions, or distraction could occur. %
%
On the other hand, having multiple sessions over a week rather than having three consecutive sessions on the same day 
also decreased internal validity due to the possibility of some students disclosing the purpose of the study to their peers despite being instructed not to do so.

\subsubsection{Timing of the tasks} \label{sec:tasks_timing}


In the original experiment, the students were given 20 minutes for the \pp tasks, 10 minutes for the solo task, 10 minutes for the first questionnaire, and 15 minutes for the second and third questionnaires. %
In the replication, the students were given 15 minutes for the in-pair tasks, 10 minutes for the solo task, 10 minutes for the first questionnaire, and 10 minutes for the second and third questionnaires,  due to the constraints imposed by their busy schedule. %


We think that the shortened duration of the in-pair tasks and the second and third questionnaires may have compromised construct validity by reducing the time span for measuring the response variables, the interaction time for assessing the partners’ skills, and the reflection time before answering each response item. Moreover, it may have weakened the effect of the treatment over confounding variables, thus decreasing also internal validity.

\subsubsection{Gendered avatars} \label{sec:gendered_avatars}

\begin{figure}%
  \centering
  \subfigure[Original experiment]{
    \includegraphics[scale=0.45]{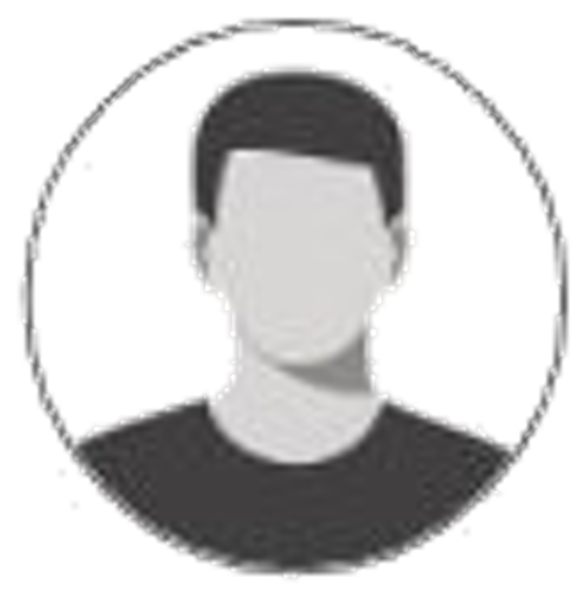}
    \includegraphics[scale=0.45]{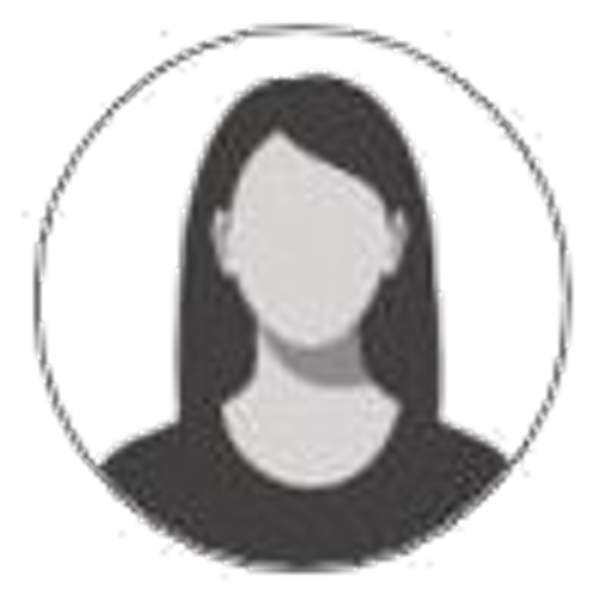}    
    \label{fig:avatars_seville}
  }\qquad\qquad
  \subfigure[Replication]{
    \includegraphics[scale=0.50]{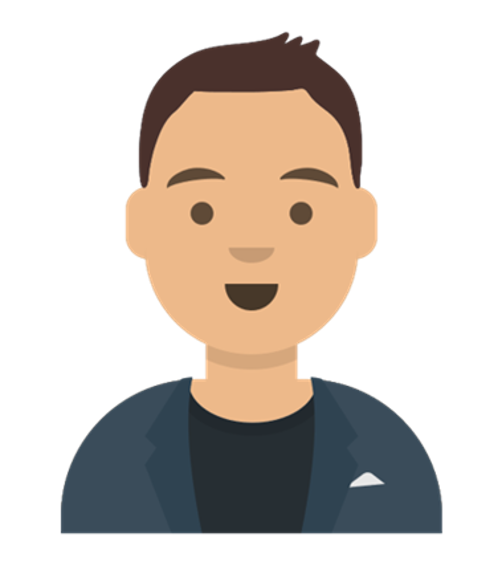}
    \includegraphics[scale=0.50]{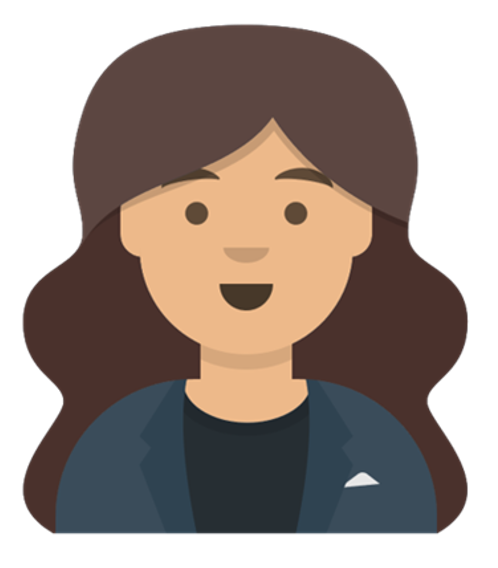}    
    \label{fig:avatars_berkeley}
  }
  \caption{Gendered avatars used in the original experiment and the replication}%
  \label{fig:avatars}%
\end{figure}

In the original experiment, the gendered avatars used in the chat windows of the subjects in the experimental group were the silhouettes shown in \figurename~\ref{fig:avatars_seville}, whereas in the replication the avatars were those shown in \figurename~\ref{fig:avatars_berkeley}, which were generated at \url{https://getavataaars.com/}. %
The subjects in the replication were also shown a gendered message at the top of the chat window indicating that their partner was connected, e.g. ``Your partner (she/her) is connected'' (see \figurename~\ref{fig:twincode_ucb_experimental} and \ref{fig:twincode_ucb_control} in \appendixname~\ref{sec:app:twincode}). 

In principle, changing the gendered silhouette avatars by more explicit ones and adding a gendered message in the chat window  would have increased construct validity, but the correlation between induced gender and perceived gender in the replication worsened with respect to the original experiment (see Section~\ref{sec:correlation_UCB}). %
As a result, we consider that this change 
decreased construct validity. %

\subsubsection{Exercise assignment}

In the original experiment, the programming exercises, which had to be solved using Javascript as the programming language, were randomly assigned to the subjects from a pool of exercises of similar complexity. %
In the replication, the programming exercises, which had to be solved in Python due to the background of the participants, were organized into two blocks (A and B) that were randomly assigned to the subjects during the experiment.

In our opinion, adapting the programming language to the background of the participants should not have any impact on experimental validity, but using two blocks of exercises instead of a pool of exercises definitely improves the blocking of the related confounding variable (see Section~\ref{sec:confounding_exercises}), thus 
increasing internal validity. %


%
%
%
    %
%


\subsection{Data Analysis} \label{sec:data_analysis_UCB}

The data analysis was performed only for those subjects considered as valid according to the same criteria than in the original experiment. %
This resulted in 23 pairs, i.e. 46 valid subjects, as previously mentioned in Section~\ref{sec:participants_UCB}. 

\subsubsection{Correlation between Induced and Perceived Gender} \label{sec:correlation_UCB}


As in the original experiment, the correlation of the induced and perceived gender in both groups was analyzed to check treatment effectiveness, especially after having changed the gendered avatars and included a gendered message at the top of the chat window, as described in Section~\ref{sec:gendered_avatars}. %

As shown in \tablename~\ref{tab:contingency_UCB}, the \level{man}/\level{man} and \level{woman}/\level{woman} effectiveness was close to 40\% in the replication whereas was close to 60\% in the original experiment (see \tablename~\ref{tab:contingency} in Section~\ref{sec:correlation}). %
Although Cramer's V for \tablename~\ref{tab:contingency_UCB} showed also a \emph{large} effect (0.530), we applied the same strict criteria than in the original experiment and decided to discard those subjects in the experimental group for whom the induced and perceived gender did not match. For the subjects in the control group, we kept those who did not perceived any gendered avatar or did not remember it.
As a result, we kept 22 subjects in the control group (10 men, 12 women) but only 9 (3 men, 6 women) in the experimental group. %

\begin{table}
  \centering
  \small
  \renewcommand{\arraystretch}{1.25}  
  \begin{tabularx}{0.85\textwidth}{cCCCC}
    
    & \multicolumn{4}{c}{\thead{Perceived Gender}} \\
    
    \cline{2-5}\noalign{\smallskip}
    
    \thead{Induced Gender}   & 
    \thead{\variable{man}}   & 
    \thead{\variable{woman}} & 
    \thead{\variable{none}}  & 
    \thead{\variable{idr}}   \\
    
    \noalign{\smallskip}\hline\noalign{\smallskip}   

    \thead[r]{\variable{man}}   & \textbf{10} (43.48\%) & 5 (21.74\%)          & 2 (8.70\%)            &  6 (26.09\%) \\
    \thead[r]{\variable{woman}} & 5 (21.74\%)           & \textbf{9} (39.13\%) & 2 (8.70\%)            &  7 (30.43\%) \\
    \thead[r]{\variable{none}}  & 0 (0.00\%)            & 1 (2.17\%)           & \textbf{30} (65.22\%) & 15 (32.61\%) \\

		\hline\noalign{\smallskip}
  \end{tabularx}
  \caption{Contingency table for induced partner's gender (\variable{ipgender}) vs. perceived partner's gender (\variable{ppgender}) in the replication} 
  \label{tab:contingency_UCB}
\end{table}

\subsubsection{Between-groups Analysis} \label{sec:between_UCB}

\begin{figure}
	\centering
		\includegraphics[scale=0.385]{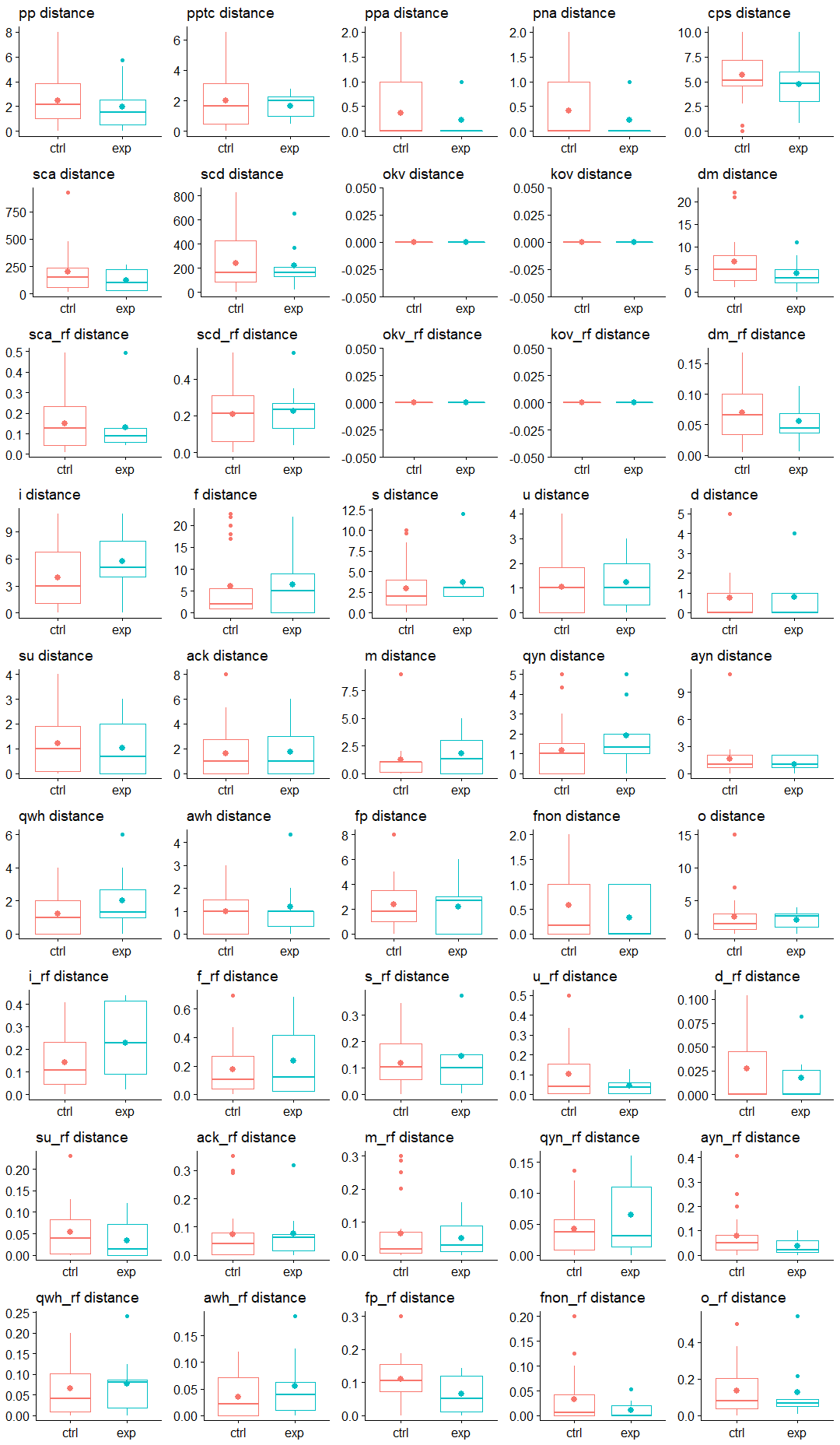}
	\caption{Boxplots of the 45 response variables for between-groups analysis in the replication}
	\label{fig:boxplot_berkeley_between}
\end{figure}

As in the original experiment, and contrary to our research hypothesis, no significant differences were observed at $\alpha$=0.05 between the control and experimental groups in the replication for any of the 45 response variables\footnote{Actually, only 41 variables were analyzed in the replication due to technical problems with the Python server used by the \twincode platform. As a result, \variable{okv}, \variable{okv\_rf}, \variable{kov}, and \variable{kov\_rf} could not be measured and, therefore, analyzed. As can be seen in \figurename~\ref{fig:boxplot_berkeley_between} and \ref{fig:boxplot_berkeley_within}, their means are 0 in all cases.} described in Section~\ref{sec:dependent_variables}, including \variable{cps}. %
The corresponding boxplots are depicted in \figurename~\ref{fig:boxplot_berkeley_between}. 

\subsubsection{Within-groups Analysis} \label{sec:within_UCB}

\begin{figure}
	\centering
		\includegraphics[scale=0.385]{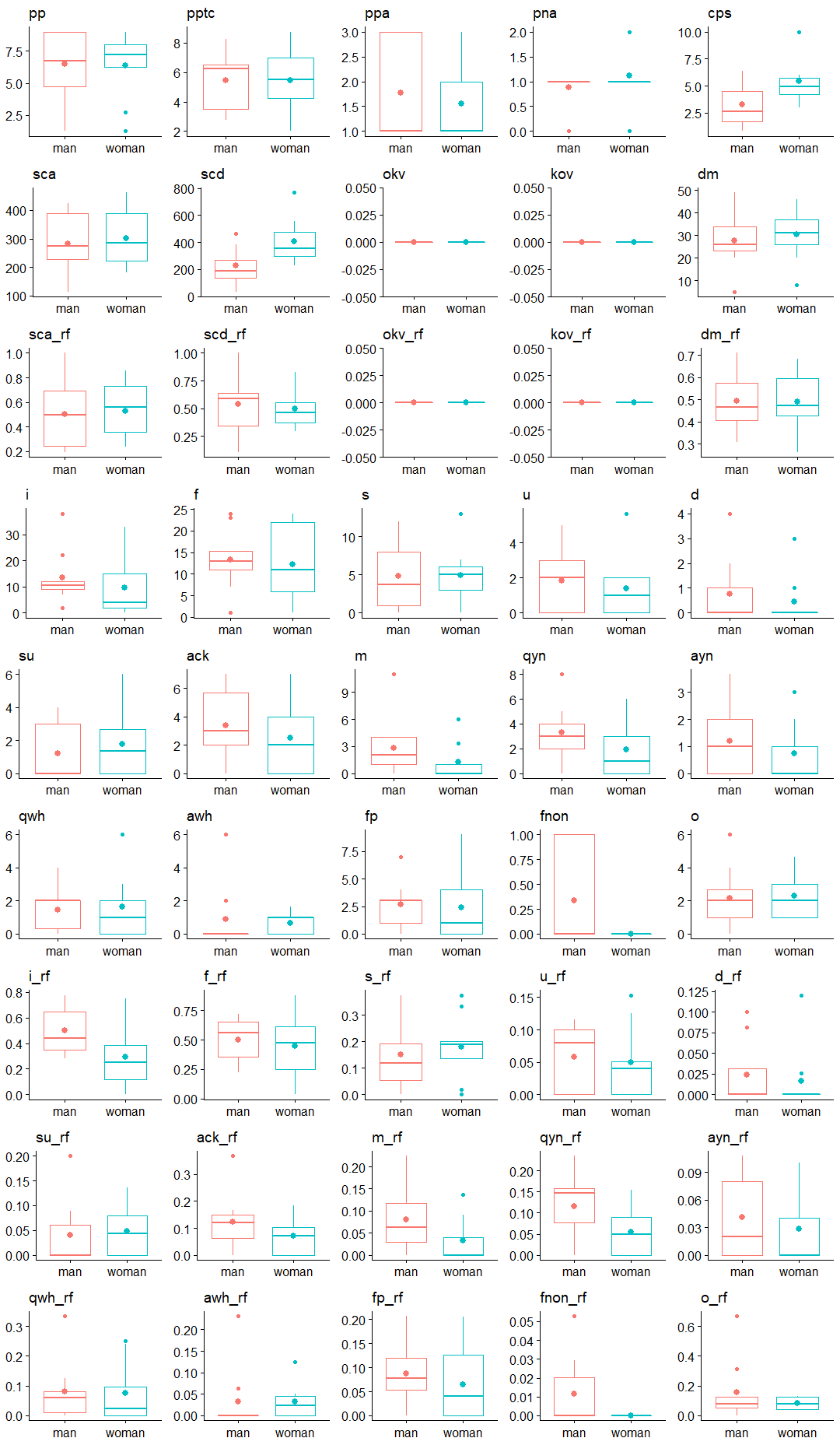}
	\caption{Boxplots of the 45 response variables for within-groups analysis in the replication}
	\label{fig:boxplot_berkeley_within}
\end{figure}

Within the experimental group (see \figurename~\ref{fig:boxplot_berkeley_within} for the corresponding boxplots), we performed the same analysis than in the original experiment, finding statistically significant differences at $\alpha$=0.05 in the following four response variables when using the perceived partner's gender (\variable{ppgender}) as a within-subjects variable. %
The four variables passed the Shapiro-Wilk normality test and were therefore analyzed using a two-sided paired t-test. Their effect sizes were computed using Cohen's $d$.

\begin{itemize}

\item 
\textbf{\variable{scd}} (source code deletions): the test detected ($p=0.0485$) that subjects deleted more source characters when they perceived their partners as a \level{woman}, with a \emph{moderate} effect size ($d=-0.775$).\newline

\item
\textbf{\variable{i\_rf}} (relative frequency of informal messages): the test detected ($p=0.0138$) that subjects increased the relative frequency of informal messages when they perceived their partners as a \level{man}, with a \emph{large} effect size ($d=1.050$).\newline

\item
\textbf{\variable{m\_rf}} (relative frequency of meta-comments or reflections): the test detected ($p=0.0377$) that subjects increased the relative frequency of meta-comments or reflections when they perceived their partners as a \level{man}, with a \emph{large} effect size ($d=0.829$).\newline

\item
\textbf{\variable{qyn\_rf}} (relative frequency of yes/no questions): the test detected ($p=0.0297$) that subjects increased the relative frequency of yes/no questions when they perceived their partners as a \level{man}, with a \emph{large} effect size ($d=0.880$).

\end{itemize}

Note that these results must be considered carefully because of the small number of selected subjects ($n$=9), and because when false discovery rate (FDR) adjustments are applied \citep{FDR-B-Y}, only the hypothesis test corresponding to the \variable{i\_rf} variable remains significant. %

No significant interactions were detected between the perceived partner's gender and the subject's gender %
for the same response variables than in the original study.



\section{Discussion and Threats to Validity} \label{sec:discussion} \label{sec:threats}

In this section, 
the original study and its external replication are discussed. Since the main concerns are about their threats to the experimental validity regarding operationalization and sampling, the discussion is organized around these type of threats, especially those that were not previously discussed in the description of the replication changes in Sections~\ref{sec:participants_UCB} and \ref{sec:execution_UCB}. 

\subsection{Operationalization of the Cause Construct --- Treatment}

The operationalization of gender bias into a treatment is not a trivial task and, according to the obtained results, we may not have designed our treatment as adequately as we intended, thus threatening construct validity. 

Considering our experimental design, telling the subjects that they were going to collaborate with a man or a woman more explicitly could have caused in many of them the suspicion of being observed about that fact, behave unnaturally and, probably, having mentioned it unintentionally during the chat messaging, thus discovering that they were being deceived about their partner's gender and invalidating the study. %

However, although the silhouetted avatars in the original experiment (see \figurename~\ref{fig:avatars_seville}) had an effectiveness close to 60\% (see \tablename~\ref{tab:contingency}), when they were changed in the replication into what we thought were more explicitly gendered avatars (see \figurename~\ref{fig:avatars_berkeley}), their effectiveness dropped under 40\% (see \tablename~\ref{tab:contingency_UCB}). %
%
%
Apart from the change of the avatars, this decrease in treatment effectiveness could have been probably affected by other factors, such as the remote setting, which increased the likelihood of distractions compared to a controlled environment such as a laboratory session, as commented in Section~\ref{sec:location_change}. Other factors could have been the reduced duration of the in-pair tasks and the second and third questionnaires, as previously discussed in Section ~\ref{sec:tasks_timing}, and the so-called \emph{Zoom burnout} \citep{zoom_burnout}, i.e., the fatigue and exhaustion caused by prolonged use of video conferencing platforms during the COVID--19 pandemic, which may have influenced the motivation and performance of students at \ucb, who are also exposed to very high levels of stress \citep{ucb_stress_2016,ucb_stress_2023}.

As commented in Section~\ref{sec:chatbots}, we are evaluating the use of chatbots together with a within-subjects design in future replications to improve the treatment and thus mitigate this threat to construct validity. %

\subsection{Operationalization of the Effect Construct --- Metrics} \label{sec:metrics}


The main goal of our work is exploring the effects of gender bias in remote \pp. Due to this exploratory nature, we have applied methodological triangulation \citep{Denzin}, observing the phenomenon from as many points of view as possible, 
with an operationalization based in 45 response variables of different types which were measured during a reasonable interaction time. 

Having said that, during the coding of the chat utterances, some of the authors who are in their fifties at the moment of writing this article perceived strong differences in how the subjects, who are Generation Z youngsters \citep{GenerationZ}, communicate compared to the way we did when we were their age. %
With all due caution, and taking into account the strong socio-political environment in Spain and the U.S. against any type of gender discrimination, we think it is possible that the presence of gender bias in people of our generation (Generation X) may have decreased two generations later, although we do not have enough evidence to affirm it. In addition, if gender bias persists, it is possible that most subjects self-censor, thus hindering the detection of its effects. %
To improve this situation, we are currently evolving the \twincode platform to include more metrics, and we are also considering the inclusion of qualitative research that might lead to new findings in future replications by widening the spectrum of collected information. %



\subsection{Sampling the Population --- Participants}

\subsubsection{Low Percentage of Women in the Original Study}

Unfortunately, the small proportion of women in STEM studies is a common issue in most higher education institutions \citep{AAUW_STEM,STEM_WOMEN}. %
The low number of women participants in the original study was an obstacle to study whether gender bias was mainly a masculine trait or if it was also present in women in any way. %
Nevertheless, the percentage of women increased substantially in the first replication without significant findings on the interaction of subject's gender with other factors. %

\subsubsection{Small Size of the Sample in the Replication}

The small size of the sample in the replication and the low effectiveness of the treatment supposed a clear threat to conclusion validity that can only be mitigated by taking the outcomes as provisional and performing more replications with bigger samples and alternative experimental designs in the future.

\subsubsection{Using Students as Subjects}

Although in other empirical studies in which subjects are \SE students, findings can be reasonably generalized to a wider community because %
the experimental tasks do not usually require high levels of industrial experience \citep{PorterBuilding},
and the students, who are the next generation of professionals, are close to the population under study \citep{KitchenhamPrelim,Runeson,Falessi}, %
the intergenerational differences commented in Section~\ref{sec:metrics} and the lack of conclusive results makes that very difficult in our case. %

%
%
%
%
%



\section{Conclusions and Future Work} \label{sec:conclusions}

After performing the original study and an external replication, we can conclude that we did not observe any effect of the gender bias treatment, nor any interaction between the perceived partner's gender and subject's gender, in any of the 45 response variables in the original study. %

With respect to the external replication, we only observed statistically significant effects within the experimental group, i.e. comparing how subjects acted when they thought their partner was a man or a woman, in four of the 45 dependent variables. %
One variable was related with changes in the behavior (source code deletions), and the other three were related with the relative frequency of different type of chat utterances (informal messages, reflections, and yes/no questions). %
In the case of the source code deletions, subjects deleted more characters 
when they perceived their partners as a woman, but the relative frequency of informal messages, reflections, and yes/no questions was higher when they perceived their partners as a man. %
We also observed a lower effectiveness of the treatment in the replication, that could be caused by the changes in the gendered avatars but also for having used a remote setting instead of a controlled environment like a laboratory session, free of distractions and interruptions. %
%
That lower effectiveness of the treatment led to a small number of selected subjects in the experimental group, thus leading to consider the replication results carefully because of the small sample they are based on, and because when FDR adjustments are applied, only the result of the relative frequency of informal messages remains significant. %

These outcomes have raised a number of potential research questions that we plan to address in the future and that are briefly described below.

\subsection{Replication in Different Cultural Background} \label{sec:cultural_differences}

The cultural differences between Spanish and U.S. students could have also influenced the outcomes of both studies, so we would like to replicate it other countries and analyze those potential differences caused by cultural backgrounds. %

\subsection{Using Chatbots as Partners and AI-based Utterance Coding} \label{sec:chatbots}

Another two research lines we would like to explore in the future are the use of chatbots as \pp partners and the use of deep learning to automatically code chat utterances, thus reducing the manual effort of carrying out a replication. %

Inspired by current trends in Psychology \citep{Bendig2019,Greer2019} and taking into account not only the absence of significant differences between groups in the original study and the replication, 
but also the difficulties in recruiting a relevant number of subjects, we are considering the possibility of changing from a between-groups design to a within-subject 
design in which each subject performs the \pp tasks with a chatbot simulating being a man or a woman instead of with another human subject. %
Obviously, developing such a chatbot is not a trivial task, but current advances in the area, such as LaMDA \citep{LaMDA}, BERT \citep{BERT}, or GPT-3 \citep{GPT-3}, make this approach a technical challenge worth exploring. %
A very relevant aspect in the development of such a chatbot is avoiding gender bias in the training data, as recently studied by \cite{McAuliffe2022}.

On the other hand, now that we have a relevant number of coded chat utterances in Spanish and English, we could use that labeled dataset to fine train a large language model system similar to those used in chatbots to classify user intents and apply it for the automatic coding of chat utterances, which is one of the most time-consuming tasks we have had to perform as experimenters in our exploratory study. %
If the results of such a fine trained system were accurate, future replications would required much less effort than the two presented in this article and experimenter bias would be considerably mitigated. %

\section*{Datasets}

The datasets generated and analyzed during the current study are available in the Zenodo repository, \url{https://doi.org/10.5281/zenodo.6783717}.

\section*{Compliance with Ethical Standards}

The authors declared that they have no conflict of interest with any aspect of the reported studies. %

The experiment protocols were approved by the Institutional Review Board (IRB) at \ucb. %
At the \us, only studies involving experimentation animals or biomedical experiments involving humans need to be approved by the Ethics Committee on Experimentation, so no approval was required in this case. 

\section*{Acknowledgements}

We would like to thank the students who volunteered to participate in the pilot studies, the original experiment and the first replication at the Universities of Seville (US) and California Berkeley (UCB). %
  We also want to thank David Brincau (undergraduate student at US) for their support in the development of the \twincode platform; %
  José Sandoval (Master's student at US) for developing \tagachat, the collaborative tool for tagging chat utterances; %
  and Daewon Kwon and Karim el Refai (undergraduate students at UCB) for their support in the evolutive changes to the \twincode platform and in the experiment execution at UCB. %
  We particularly acknowledge Vron Vance (UCB alumnus, Data Analyst at Google) for their assistance regarding inclusive language around gender identity. %
	Last but not least, we would like to thank the anonymous reviewers for their valuable comments and suggestions that helped us improve the quality and clarity of this article. \\\\%
  \FUNDING

\bibliographystyle{sn-basic}
\bibliography{duran_emse_twincode_2022}

\clearpage

\appendix


\section{Questionnaire \#1 and \#2 response items} \label{sec:app:response_items}

In this section, the response items of the scales used in questionnaires \#1 and \#2 are enumerated. %
Those scales were analyzed for internal consistency using the data collected during the pilot studies, and the results of those analysis consisting in the Pearson's correlations, Cronbach's $\alpha$, and principal components scree plot are also reported \citep{UCLA_Cronbach}, indicating whether some response items were dropped or not according to the obtained results. %

\subsection{Response items for perceived productivity scale (\variable{pp})} \label{sec:app:pp}

All the items in this questionnaire section, entitled as ``\variable{Solo programming or pair programming?}'', are 0--10 numerical response items in which 0 means ``\variable{programming solo}'', 5 means ``\variable{the same in both cases}'', 10 means ``\variable{programming in pairs}''. \newline%

\noindent\begin{tabularx}{\textwidth}{lX}

\textbf{\variable{pp}$_1$} &
Regarding the programming exercises you just did, how do you think you would have been \textbf{more productive}, programming solo or programming with the partner assigned to you?\newline\\
 
\textbf{\variable{pp}$_2$} &
Regarding the programming exercises you just did, how do you think you would have achieved a \textbf{better program quality}, programming solo or programming with the partner assigned to you?\newline\\

\textbf{\variable{pp}$_3$} &
Regarding the programming exercises you just did, how do you think you would have developed a \textbf{more reliable} program, i.e., a program more likely to run without failures, programming solo or programming with the partner assigned to you?\newline\\

\textbf{\variable{pp}$_4$} &
Regarding the programming exercises you just did, how do you think you would have \textbf{enjoyed more}, programming solo or programming with the partner assigned to you?

\end{tabularx}

\begin{figure}[h]
	\centering
    \includegraphics[scale=0.375,trim={0.75cm 0.75cm 0.75cm 1.50cm},clip]{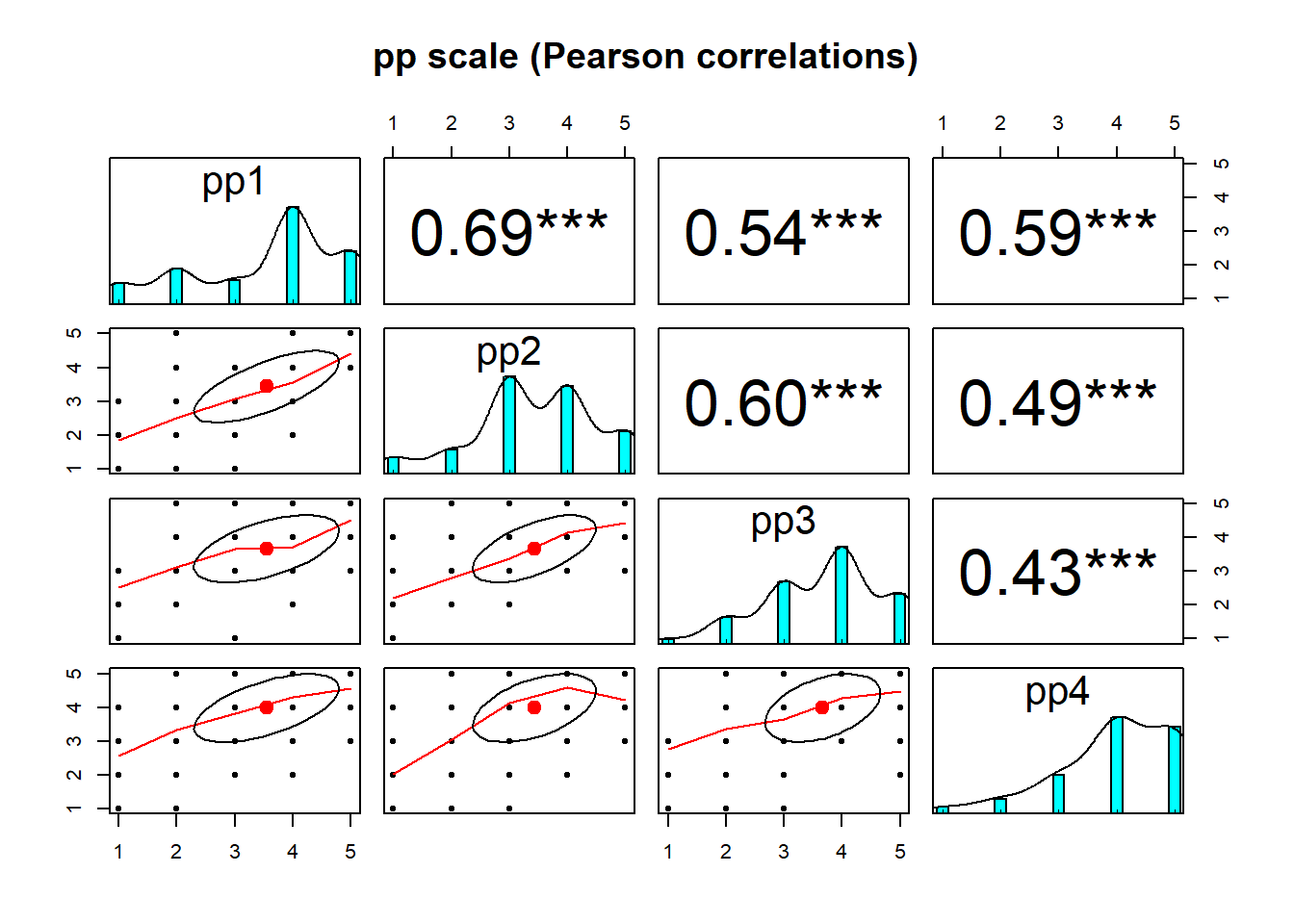}
    \quad
    \includegraphics[scale=0.325,trim={0.75cm 1.5cm 0.75cm 1.65cm},clip]{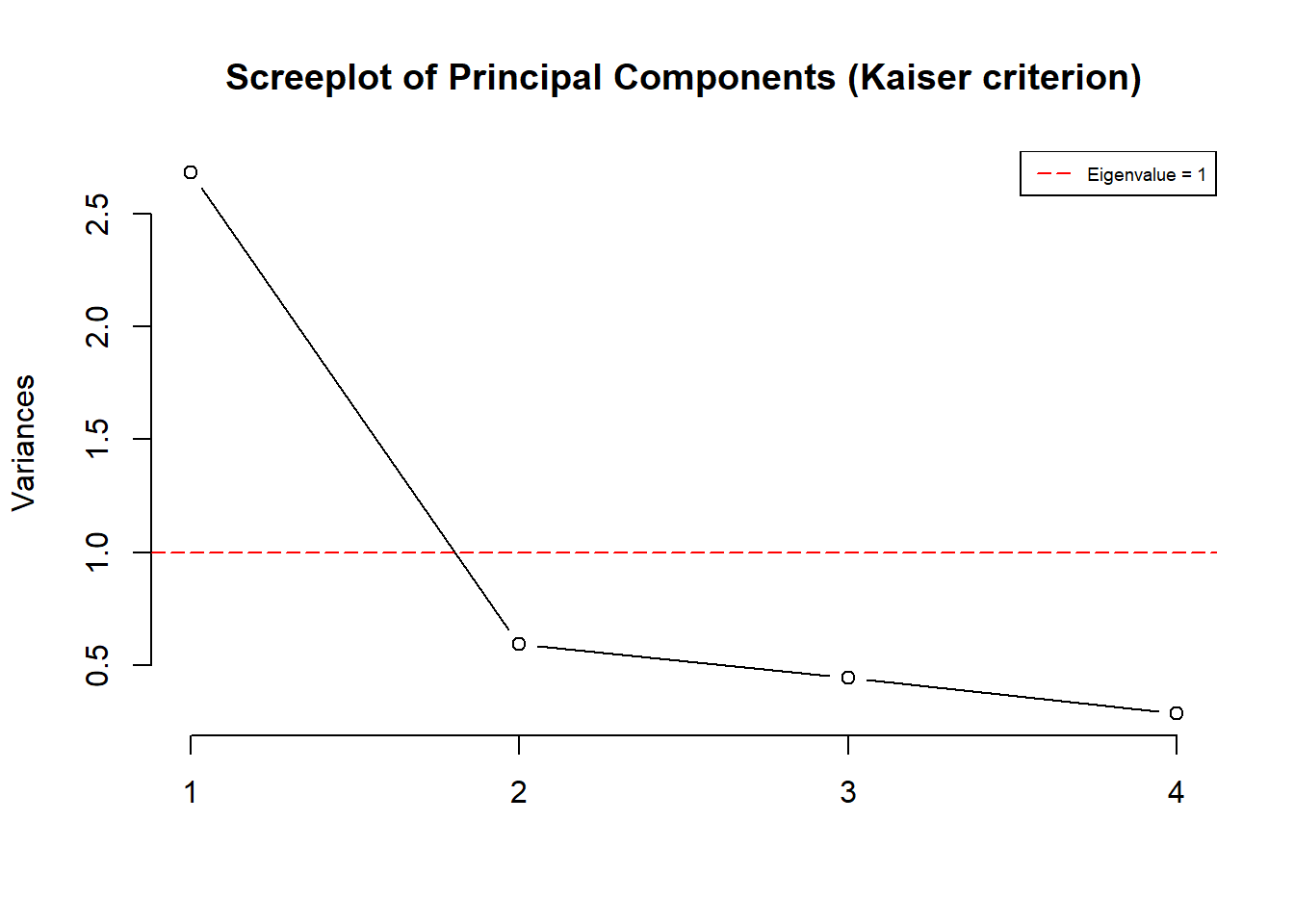}
	\caption{Pearson correlations and scree plot of \variable{pp} scale items}
	\label{fig:pp_correlations}
\end{figure}

As shown in \figurename~\ref{fig:pp_correlations}, all the items presented high Pearson correlations with Cronbach's $\alpha = 0.83$, and the scree plot confirmed they were unidimensional according to the Kaiser criterion. As a result, all of them were kept after the reliability analysis on the data from the pilot studies. %

\subsection{Response items for partner’s perceived technical competency (\variable{pptc})} \label{sec:app:pptc}

All the items in this questionnaire section, entitled as ``\variable{My partner or me?}'', are 0--10 numerical response items in which 0 means ``\variable{me}'', 5 means ``\variable{both equally}'', 10 means ``\variable{my partner}''.\\

\noindent\begin{tabularx}{\textwidth}{lX}

\textbf{\variable{pptc}$_1$} &
During the programming exercises you just did, who do you think had more \textbf{knowledge and technical skills}, you or the partner assigned to you?\newline\\

\textbf{\variable{pptc}$_2$} &
During the programming exercises you just did, who do you think has been more \textbf{cooperative}, you or the partner assigned to you?\newline\\

\textbf{\variable{pptc}$_3$} &
During the programming exercises you just did, who do you think has had a \textbf{faster pace at solving the exercises}, you or the partner assigned to you?\newline\\

\textbf{\variable{pptc}$_4$} &
During the programming exercises you just did, who do you think has \textbf{led more to the solutions}, you or the partner assigned to you?

\end{tabularx}

\begin{figure}[h]
	\centering
    \includegraphics[scale=0.375,trim={0.75cm 0.75cm 0.75cm 1.50cm},clip]{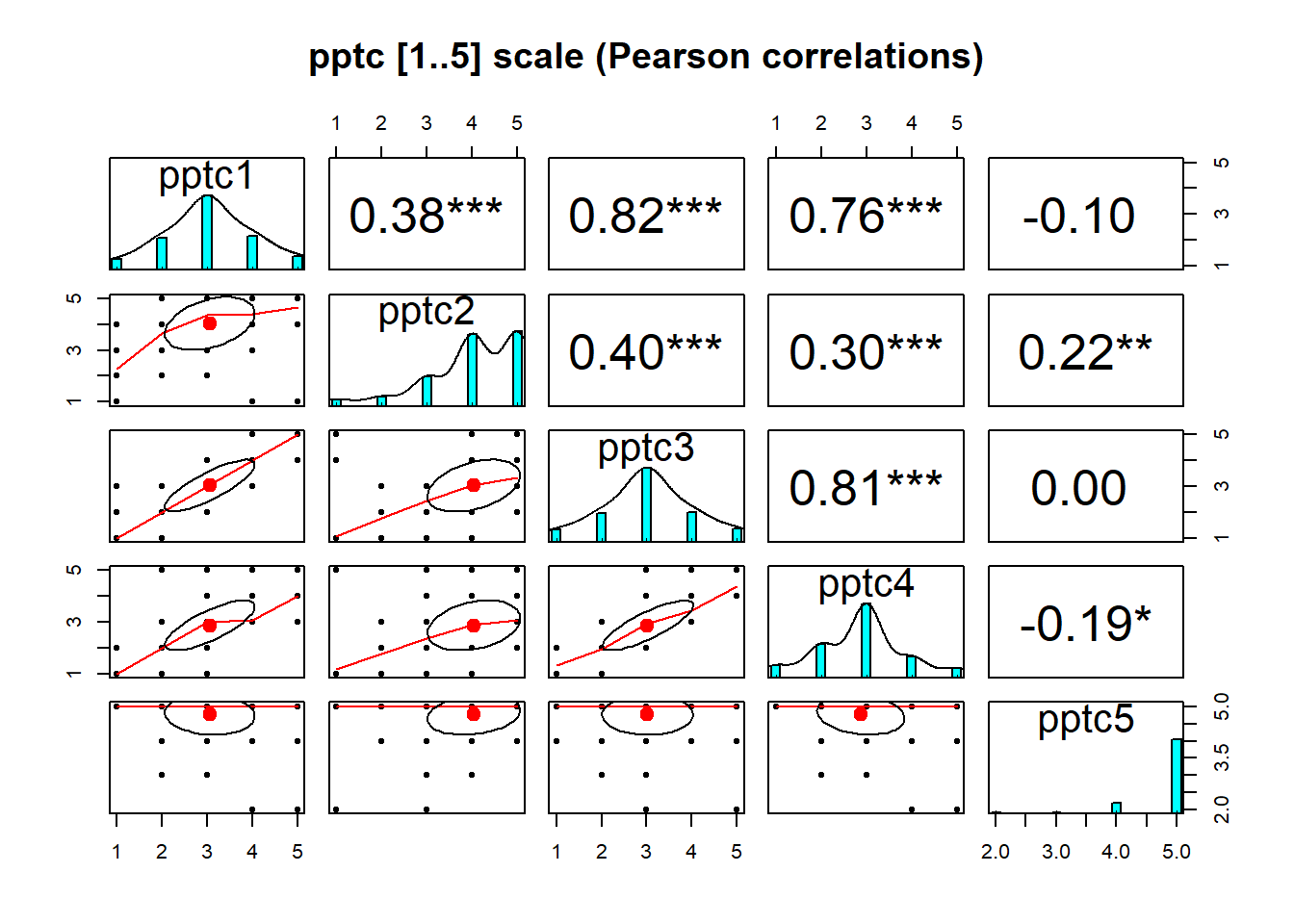}
    \quad
    \includegraphics[scale=0.325,trim={0.75cm 1.5cm 0.75cm 1.65cm},clip]{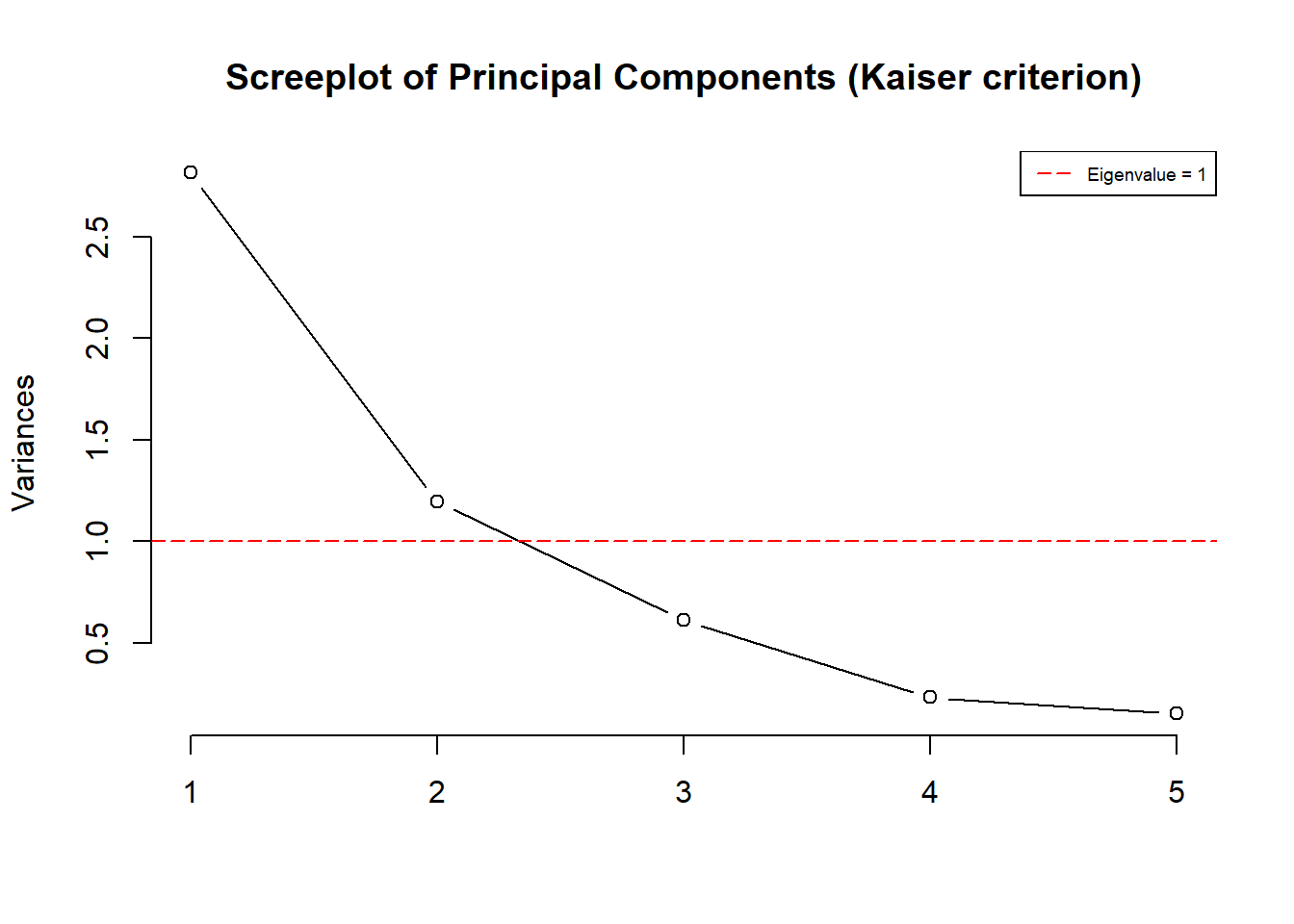}
	\caption{Pearson correlations and scree plot of the initial version of \variable{pptc} scale items}
	\label{fig:pptc_correlations_v1}
\end{figure}

\begin{figure}[h]
	\centering
    \includegraphics[scale=0.375,trim={0.75cm 0.75cm 0.75cm 1.50cm},clip]{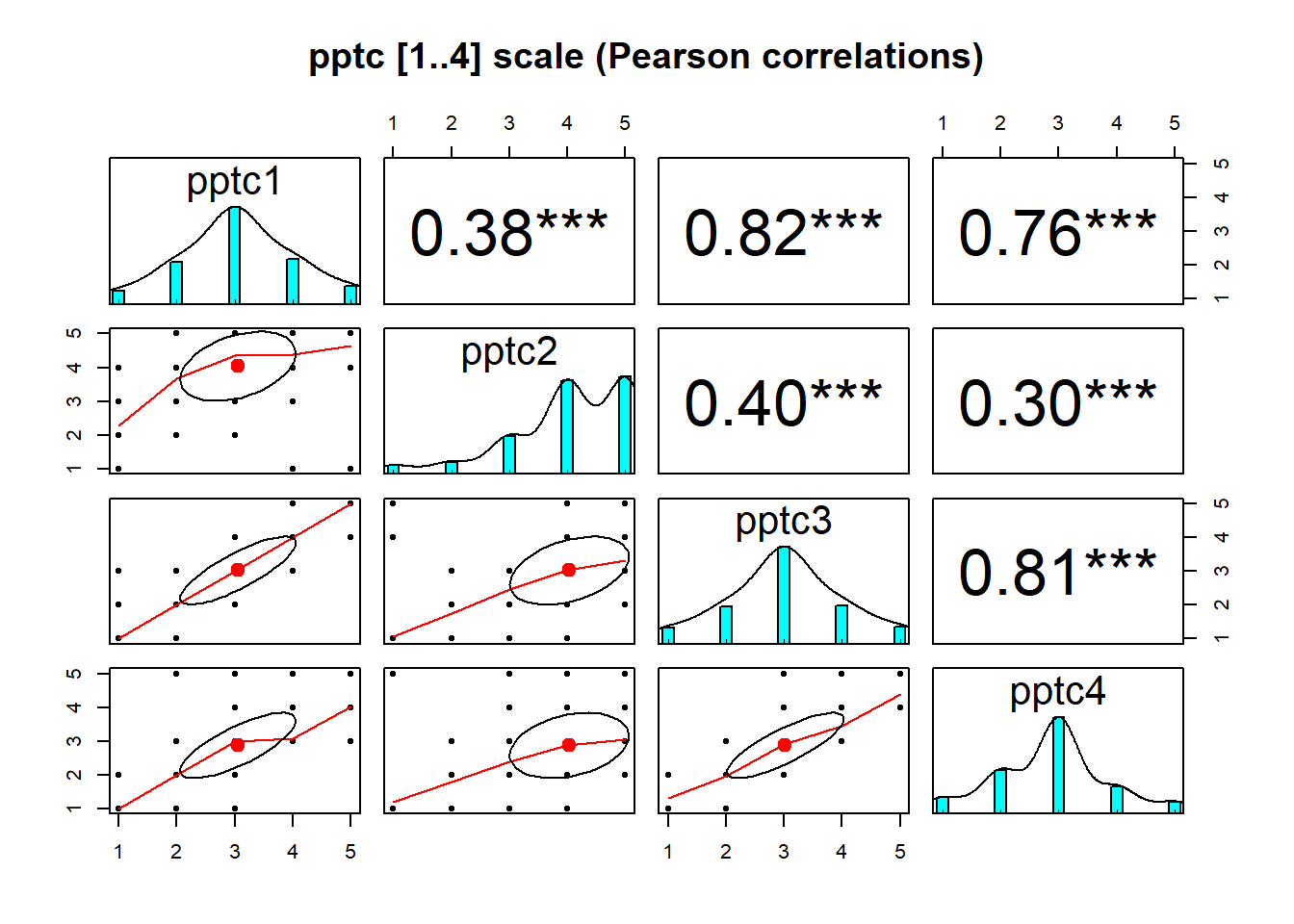}
    \quad
    \includegraphics[scale=0.325,trim={0.75cm 1.5cm 0.75cm 1.65cm},clip]{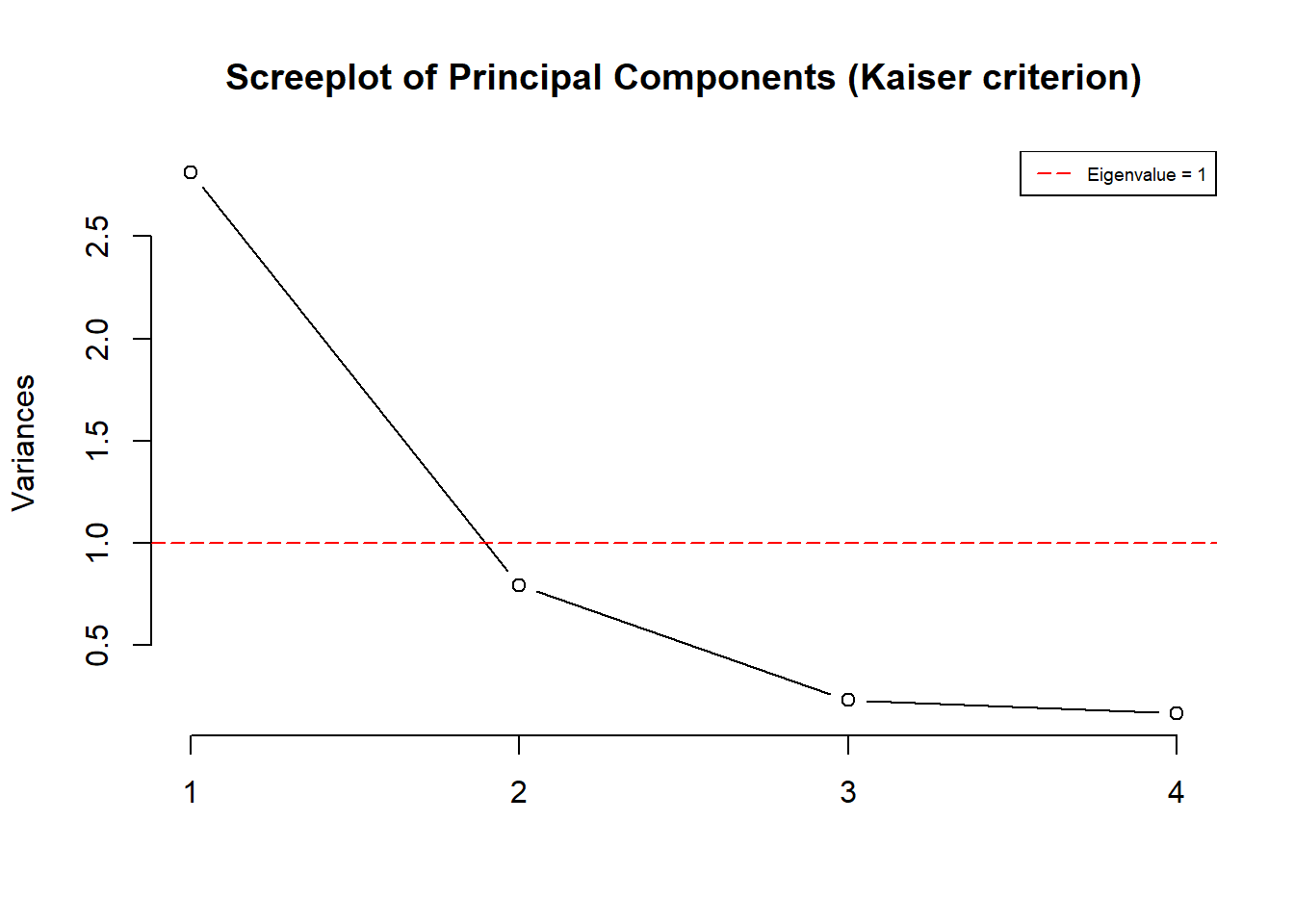}
	\caption{Pearson correlations and scree plot after dropping \variable{pptc}$_5$ from \variable{pptc} scale}
	\label{fig:pptc_correlations_v2}
\end{figure}

As shown in \figurename~\ref{fig:pptc_correlations_v1}, in the initial version of the scale used in the pilot studies, the \variable{pptc}$_5$ item, which asked whether the assigned partner had been condescending, presented low correlations with the rest of the items in the scale and the scree plot indicated two factors. After removing that uncorrelated item, the Cronbach's $\alpha$ increased from 0.73 to 0.85, and the scree plot indicated only one factor, as shown in \figurename~\ref{fig:pptc_correlations_v2}. %

\subsection{Response item for partner’s perceived positive and negative aspects (\variable{ppa} and \variable{pna})} \label{sec:app:ppna}

The only item in this questionnaire section, entitled as ``\variable{Describe your partner}'', is a free text field in which subjects are instructed to describe the most positive and most negative aspects of the partner assigned to them in the programming exercises they just did, indicating the positive ones with a "+" sign and the negative ones with a "-" sign in front of each aspect.\\

\subsection{Response items for compared partners’ skills (\variable{cps})} \label{sec:app:cps}

All the items in this questionnaire section, entitled as ``\variable{First or second partner?}'', are 0--10 numerical response items in which 0 means ``\variable{first partner}'', 5 means ``\variable{both equally}'', 10 means ``\variable{second partner}''.\\

\noindent\begin{tabularx}{\textwidth}{lX}

\textbf{\variable{cps}$_1$} &
Comparing your assigned partners in sessions 1 and 3, who do you think provided \textbf{more clear and constructive feedback}, your first partner or your second partner?\newline\\

\textbf{\variable{cps}$_2$} &
Comparing your assigned partners in sessions 1 and 3, who do you think was \textbf{easier to communicate with}, your first partner or your second partner?\newline\\  

\textbf{\variable{cps}$_3$} &
Comparing your assigned partners in sessions 1 and 3, who do you think who do you think was \textbf{more knowledgeable about the subject material}, your first partner or your second partner?\newline\\

\textbf{\variable{cps}$_4$} &
Comparing your assigned partners in sessions 1 and 3, who do you think would be a \textbf{better project partner}, your first partner or your second partner?\newline\\

\textbf{\variable{cps}$_5$} &
Comparing your assigned partners in sessions 1 and 3, who do you think would be a \textbf{better teaching assistant}, your first partner or your second partner

\end{tabularx}

\begin{figure}[h]
	\centering
    \includegraphics[scale=0.375,trim={0.75cm 0.75cm 0.75cm 1.50cm},clip]{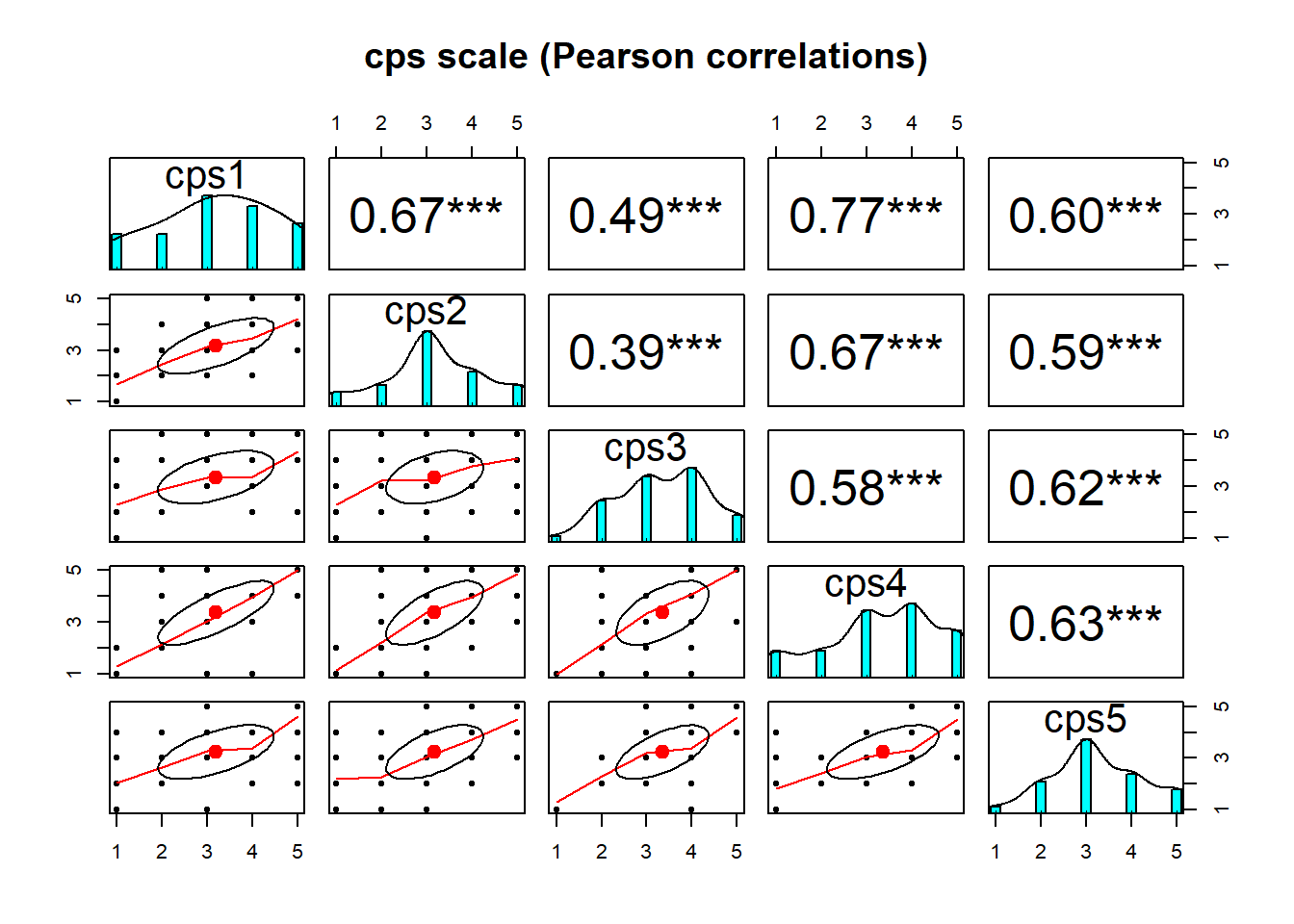}
    \quad
    \includegraphics[scale=0.325,trim={0.75cm 1.5cm 0.75cm 1.65cm},clip]{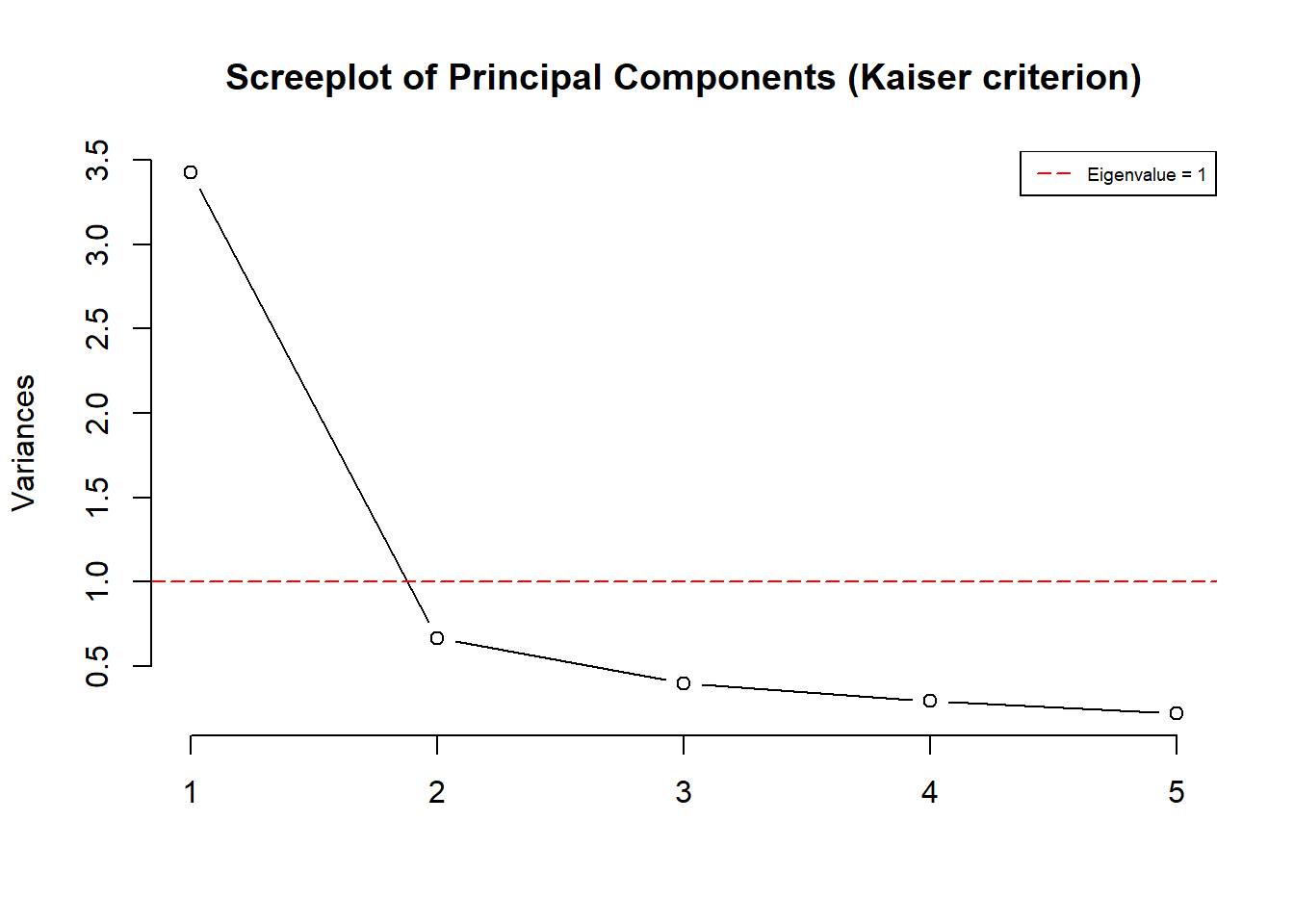}
	\caption{Pearson correlations and scree plot of \variable{cps} scale items}
	\label{fig:cps_correlations}
\end{figure}

As shown in \figurename~\ref{fig:cps_correlations}, all the items presented high Pearson correlations with Cronbach's $\alpha = 0.88$, and the scree plot confirmed they were unidimensional according to the Kaiser criterion. As a result, all of them were kept after the reliability analysis on the data from the pilot studies. %

 \clearpage


\section{Evolution of the \twincode User Interface} \label{sec:app:twincode}

\begin{figure*}[h]
	\centering
  \subfigure[Experimental group --- gendered avatar]{       
    \includegraphics[height=0.36\textheight]{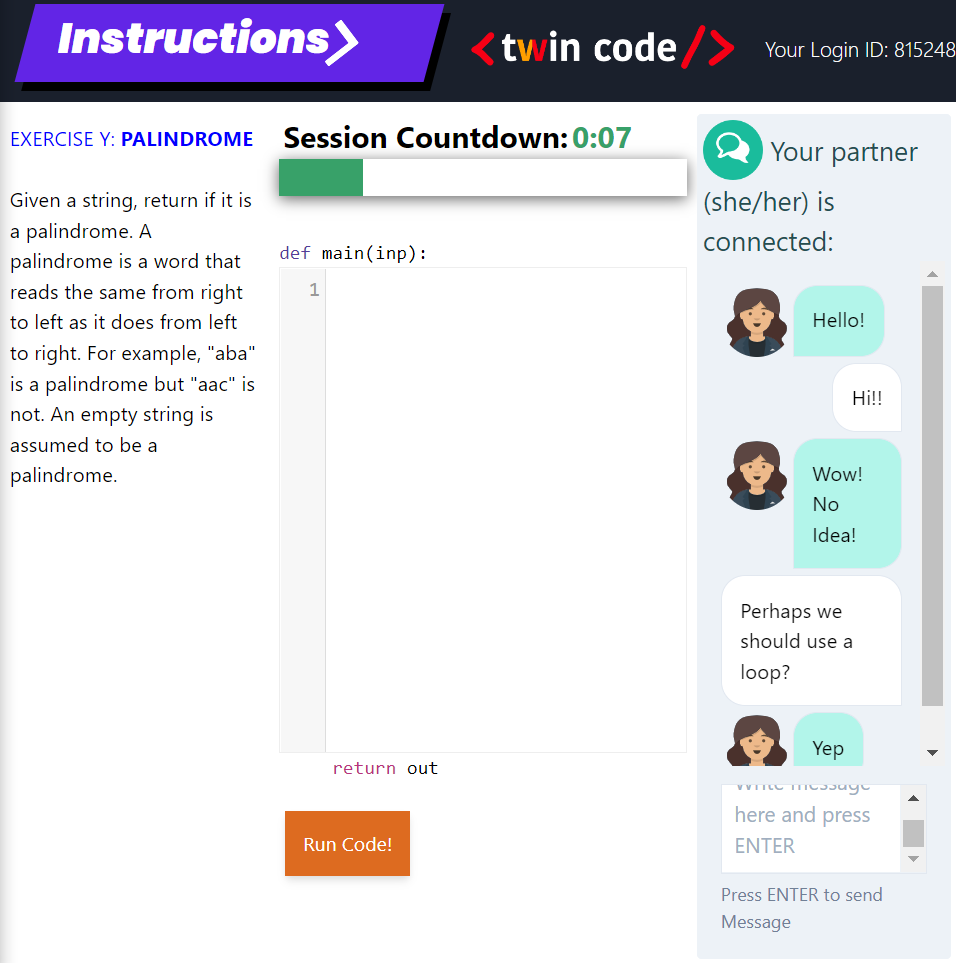}  
    \label{fig:twincode_ucb_experimental}
  }
  \subfigure[Control group --- no avatar]{
    \includegraphics[height=0.36\textheight]{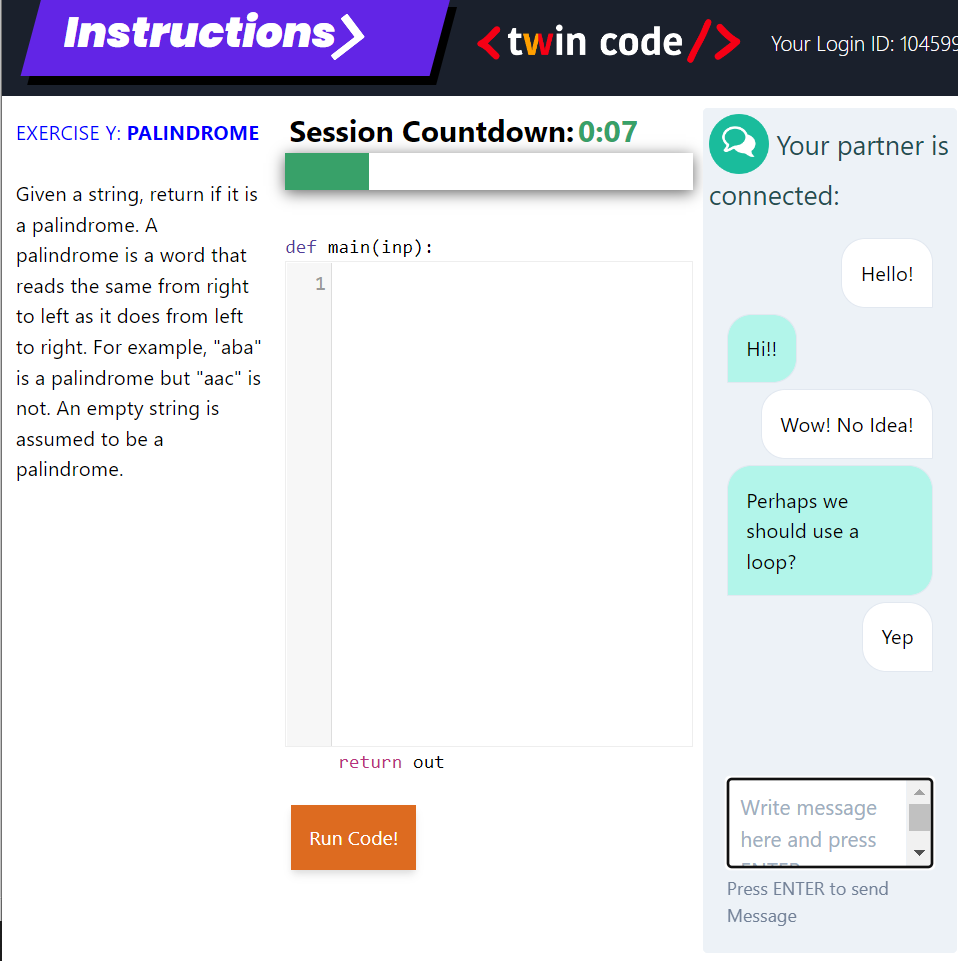}
    \label{fig:twincode_ucb_control}
  }
	\caption{\twincode user interface for subjects in the experimental and control groups (replication version)}
	\label{fig:twincode_ucb}
\end{figure*}

The \twincode user interface used in the external replication at \ucb is shown in \figurename~\ref{fig:twincode_ucb_experimental} and \ref{fig:twincode_ucb_control}. %

\clearpage

\section{User Interface of \tagachat} \label{sec:app:tagachat}

\begin{figure*}[h]
	\centering
		\includegraphics[width=0.95\textwidth]{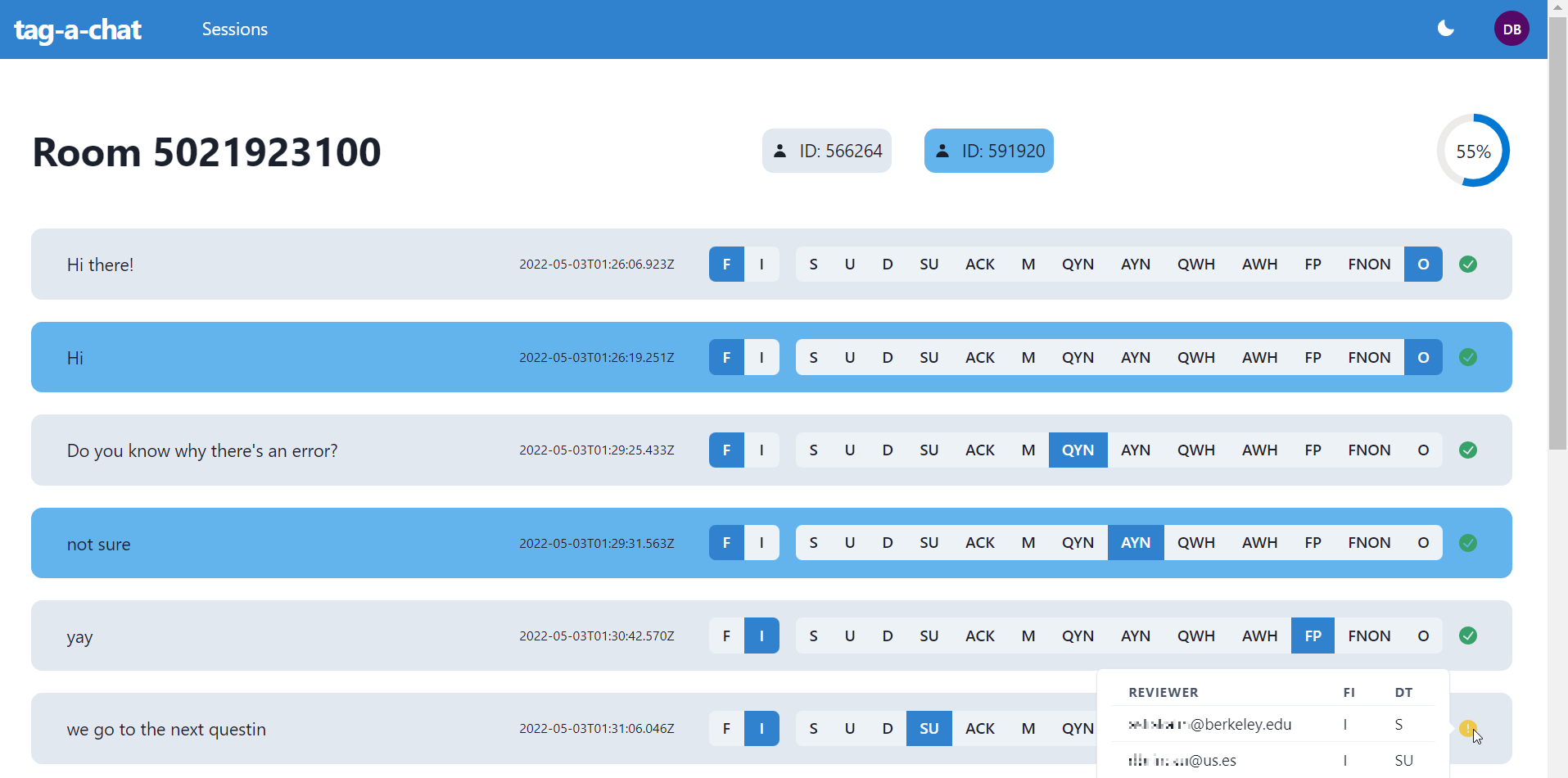}
	\caption{User interface of the \tagachat tool}
	\label{fig:tagachat}
\end{figure*}

The user interface of the \tagachat tool used for collaboratively coding chat utterances is shown in \figurename~\ref{fig:tagachat}. %

\end{document}